\documentclass{aa}
\pdfoutput=1
\usepackage{txfonts}
\usepackage{graphicx}
\usepackage{xcolor}

\def\H2{H$_2$}

\begin{document}

\title{Seeds of Life in Space (SOLIS) V.
Methanol and acetaldehyde in the protostellar jet-driven shocks L1157-B0 and B1}
\author{C. Codella \inst{1,2} 
\and C. Ceccarelli \inst{2,1} 
\and E. Bianchi\inst{2}
\and N. Balucani\inst{3,1,2}
\and L. Podio\inst{1}
\and P. Caselli\inst{4}
\and S. Feng\inst{5,6,7}
\and B. Lefloch\inst{2}
\and A. L\'opez-Sepulcre\inst{2,8}
\and R. Neri\inst{8}
\and S. Spezzano\inst{4}
\and M. De Simone\inst{2}
        }

\institute{
INAF, Osservatorio Astrofisico di Arcetri, Largo E. Fermi 5,
50125 Firenze, Italy
\and
Univ. Grenoble Alpes, CNRS, Institut de
Plan\'etologie et d'Astrophysique de Grenoble (IPAG), 38000 Grenoble, France
\and
Dipartimento di Chimica, Biologia e Biotecnologie, Via Elce di Sotto 8, 06123 Perugia, Italy
\and
Max-Planck-Institut f\"ur extraterrestrische Physik (MPE), 
Giessenbachstrasse 1, 85748 Garching, Germany
\and
National Astronomical Observatory of China, Datun Road 20, Chaoyang, Beijing, 100012, P. R. China
\and
CAS Key Laboratory of FAST, NAOC, Chinese Academy of Sciences,  P. R. China
\and
National  Astronomical  Observatory  of  Japan,  2  Chome-21-1  Osawa,  Mitaka-shi,  Tokyo-to  181-0015,  Japan
\and
Institut de Radioastronomie Millim\'etrique, 300 rue de la Piscine, Domaine
Universitaire de Grenoble, 38406, Saint-Martin d'H\`eres, France
}

\offprints{C. Codella, \email{codella@arcetri.astro.it}}
\date{Received date; accepted date}

\authorrunning{Codella et al.}
\titlerunning{Methanol, acetaldehyde, and complex organics in L1157-B0 and B1}

\abstract
{It is nowadays clear that a rich organic chemistry takes place in
  protostellar regions. However, the processes responsible for it,
  that is, the dominant formation routes to interstellar complex organic
  molecules, are still a source of debate. Two paradigms have been
  evoked: the formation of these molecules on interstellar dust
  mantles and their formation in the gas phase from simpler species
  previously synthesised on the dust mantles.}
{In the past, observations of protostellar shocks have been used to
  set constraints on the formation route of formamide (NH$_2$CHO), exploiting its
  observed spatial distribution and comparison with astrochemical
  model predictions. In this work, we follow the same strategy to
  study the case of acetaldehyde (CH$_3$CHO).}
{To this end, we used the data obtained with the IRAM-NOEMA
  interferometer in the framework of the Large Program SOLIS to image
  the B0 and B1 shocks along the L1157 blueshifted outflow in methanol
  (CH$_3$OH) and acetaldehyde line emission.}
{We imaged six CH$_3$OH and eight CH$_3$CHO lines which cover upper-level energies up to $\sim$ 30 K. Both species trace the B0
  molecular cavity as well as the northern B1 portion, that is, the
  regions where the youngest shocks ($\sim$ 1000 yr) occurred. The
  CH$_3$OH and CH$_3$CHO emission peaks towards the B1b clump, where
  we measured the following column densities and relative abundances:
  1.3 $\times$ 10$^{16}$ cm$^{-2}$ and 6.5 $\times$ 10$^{-6}$
  (methanol), and 7 $\times$ 10$^{13}$ cm$^{-2}$ and 3.5 $\times$
  10$^{-8}$ (acetaldehyde). We carried out a non-LTE (non-Local
Thermodinamic Equilibrium) Large Velocity Gradient (LVG) analysis of
  the observed CH$_3$OH line: the average kinetic temperature and
  density of the emitting gas are $T_{\rm kin}$ $\sim$ 90 K and
  $n_{\rm H_2}$ $\sim$ 4 $\times$ 10$^{5}$ cm$^{-3}$, respectively.
  The CH$_3$OH and CH$_3$CHO abundance ratio towards B1b is 190,
  varying by less than a factor three throughout the whole B0--B1
  structure.}
{Comparison of astrochemical model predictions with the observed
  methanol and acetaldehyde spatial distribution does not allow us to
  distinguish whether acetaldehyde is formed on the grain mantles or
   in the gas phase, as its gas-phase formation, which is dominated by
  the reaction of ethyl radical (CH$_3$CH$_2$) with atomic oxygen, is
  very fast.  Observations of acetaldehyde in younger shocks,
  for example those of $\sim$ 10$^{2}$ yr old, and/or of the ethyl radical, whose
  frequencies are not presently available, are necessary to settle the
  issue.}


\maketitle

\section{Introduction}

One of the open questions in astrochemistry refers to how chemistry richness
develops and evolves in star forming regions. More specifically, the molecular complexity during the evolutionary stages that lead a
molecular cloud to form a solar-type star and its planetary system has yet to be elucidated.
Particularly relevant are the abundances of the so-called
interstellar complex organic molecules (iCOMs: molecules with at least
six atoms: Herbst \& van Dishoeck 2009; Ceccarelli et al. 2017), as they
may be considered as small blocks from which complex pre-biotic species
can build up (e.g. Caselli \& Ceccarelli 2012; Belloche et al. 2014).
The present consensus is that after an initial cold phase in which
icy mantles enriched with hydrogenated species form on the dust grains
(e.g. Tielens \& Hagen 1980; Watanabe \& Kouchi 2002; Rimola et
al. 2014), the iCOMs synthesis can either occur on the grain mantle
surfaces initiated by energetic processing (e.g. Garrod \& Herbst
2006; Garrod et al. 2008; Ruaud et al. 2015; Vasyunin et al. 2017) or
in the gas phase initiated by the sublimation of the mantles
themselves (e.g. Charnley et al. 1992; Balucani et al. 2015; Codella
et al. 2017; Skouteris et al. 2018).
In both cases, the composition of the grain mantles therefore plays a
paramount role in the gaseous, observable abundance of iCOMs (see e.g. Burkhardt et al. 2019) -- in the
former possible scenario, because the mantle species are directly injected from the
mantles into the gas; and in the latter possible scenario, because they feed gas
reactions leading to iCOMs.

In this context, protostellar shocks are particularly
precious because they can provide not only the observed iCOM
abundances but also the approximate times at which they are formed. Indeed, at the
passage of the shock, molecules present in the grain mantles are
injected into the gas phase either because of sputtering (gas-grain
collisions) or shattering (grain--grain) processes (e.g. Flower \&
Pineau Des For\^ets 1994; Caselli et al. 1997; Schilke et al. 1997;
Gusdorf et al. 2008). Once in the gas phase, the injected mantle
species undergo reactions that destroy and/or form iCOMs (unless
multiple shocks occur at the very same place, which is unlikely):
therefore, the gaseous iCOM abundances are intrinsically
time dependent. Fortunately, the chemical evolution time, which is a few
thousand years at most, is of the order of the age of young
protostellar shocks (e.g. Herbst \& van Dishoeck 2009; Ceccarelli et al. 2017). Thus, sufficiently high spatial resolution
observations towards these objects allow us to decipher the different
time-dependent abundances of iCOMs and consequently provide extremely
strong constraints on the formation and destruction of iCOMs. For example, this method was applied
to
constrain the formation route of formamide in the jet-driven protostellar shock L1157-B1 (Codella et al. 2017:
hereinafter CC17). Previous observations have shown that this latter shock is
extremely rich in iCOMs (Arce et al. 2008; Lefloch et al. 2017), suggesting
that it may be an ideal case to look for others
with the same method as that used by Codella et al.

In this article, we continue the systematic study of the
formation and destruction routes for iCOMs in the L1157-B1 shock started in CC17.
Here we focus on the acetaldehyde (CH$_3$CHO) and methanol (CH$_3$OH). 
As in CC17, we make use of the observations from the
IRAM/NOEMA (NOrthern Extended Millimeter
Array\footnote{http://iram-institute.org/EN/noema-project.php}) Large
Project SOLIS ({\it Seeds Of Life In Space}: Ceccarelli et al. 2017),
whose goal is to investigate the formation and destruction of iCOMs
during the early stages of solar-type star formation.  The article is
organised as follows: in \S \ref{sec:l1157-b1} we describe the source
and summarise the previous studies of its  astrochemistry. In \S
\ref{sec:observations} we describe the observations, in \S
\ref{sec:results} we report the results of the analysis of these observations,
in \S \ref{sec:discussion} we discuss the impact of those results, and in
\S \ref{sec:conclusions} we present our conclusions.

\section{L1157-B1: an astrochemical  laboratory}\label{sec:l1157-b1}

The chemically rich molecular outflow driven by the L1157-mm Class 0
protostar ($d$ = 352 pc; Zucker et al. 2019) has been extensively used
to investigate how the gas chemistry is altered by the injection of
the species frozen onto the dust mantles.  L1157-mm drives an episodic
and precessing jet (Gueth et al. 1996; Podio et al. 2016), which
produced several cavities well observed using both single-dish
antennas and interferometers (e.g. Gueth et al. 1998; Lefloch et
al. 2012).  In particular, the southern blueshifted outflow is
associated with several shocked regions produced by 
impacts between jets and cavities. called B0, B1, and B2 (see also Fig. 1).  The brightest
shocked region, B1, is one of the targets of the IRAM NOEMA SOLIS
Large Program.  In particular, B1 consists of a series of shocks
caused by different episodes of ejection hitting the cavity
wall.  These shocks produced a clumpy structure (identified using
several molecular species: see e.g. Tafalla \& Bachiller 1995; Codella
et al. 2009; Lefloch et al. 2012; Benedettini et al. 2013), which can
be classified as follows: (i) a northern arch-like structure composed
of the clumps B1a-e-f-b (see Fig. 1), plus (ii) two clumps, labelled
B1c and B1d, which are the oldest ones (kinematical age $\sim$ 1400
yr, from Podio et al. 2016, corrected for the new distance of 352 pc by Zucker et al. 2019), 
being the farthest away from the source, close to the B1 apex.

Within the context of the study of iCOMs, the advantages of observing
L1157-B1 are that: (i) the observations are not polluted by the protostar
emission, as it is located at a projected distance of 0.11 pc, and (ii) several previous
studies have shown that grain-mantle species have been injected, due
to shock-induced sputtering and shattering, into
the gas phase (e.g. Bachiller et al. 2001; Arce et al. 2008; Codella
et al. 2010; Fontani et al. 2014; Lefloch et al. 2017).

In the first SOLIS paper on L1157-B1 (CC17), we compared the
spatial distribution of the acetaldehyde (CH$_3$CHO) line emission
with that from formamide (NH$_2$CHO), finding a striking chemical
stratification. This implies a NH$_2$CHO/CH$_3$COH abundance ratio
which varies within the B1 structure: specifically, the formamide line
emission is associated with older shocked gas with respect to
acetaldehyde.
Thanks to the comparison of the observations with astrochemical model
predictions, we provided evidence that formamide is
formed  in L1157-B1 by gas-phase chemical reactions. However, we were unable to
provide any indication on the formation route of acetaldehyde.

\section{Observations}\label{sec:observations}

The L1157-B0/B1 shocks were observed at 3 mm
with the IRAM NOEMA array
during two tracks in October 2016 (7 antennas), and three tracks in 
January 2017 (8 antennas) using both
the C and D configurations. The shortest and longest 
baselines are 24 m and 304 m, respectively.
These configurations allowed us 
to recover emission at scales (with a full efficiency) up to $\sim$ 14$\arcsec$.
The primary beam was 64$\arcsec$.
The phase centre is $\alpha({\rm J2000})$ = 20$^h$ 39$^m$ 10$\fs$2,
$\delta({\rm J2000})$ = +68$\degr$ 01$\arcmin$ 10$\farcs$5.
The CH$_3$OH and CH$_3$CHO
lines listed in Table 1 were observed using the WideX backends, with a total
spectral band of $\sim$ 4 GHz, and a spectral resolution of 
1.95 MHz ($\sim$ 6 km s$^{-1}$). Several CH$_3$OH lines were also
observed with 80 MHz backends with a spectral resolution of
156 kHz ($\sim$ 0.48 km s$^{-1}$).
Calibration was carried out following standard procedures
using GILDAS-CLIC\footnote{http://www.iram.fr/IRAMFR/GILDAS}.
The bandpass was calibrated on 3C454.3, while the absolute
flux was fixed by observing MWC349, 2200+420, 2013+370, and 2007+659, the latter also being
used to set the gains in phase and amplitude.
The final uncertainty on the absolute flux scale is $\leq$ 15\%.
The phase rms was $\le$ 50$\degr$,
the typical precipitable water vapour (PWV) was from 2 mm to 20 mm,
and the system temperatures were $\sim$ 50--100 K (D) and $\sim$ 50--250 K (C).
The rms noise in the 1.95 MHz channels was 4--20 mJy beam$^{-1}$,
depending on the frequency (see Table 1).
Images were produced using natural weighting, and restored with a
clean beam of 2$\farcs$97 $\times$ 2$\farcs$26 (PA=--155$\degr$).

\section{Results}\label{sec:results}

We imaged six methanol emission lines  
towards the L1157 blueshifted outflow, covering 
upper level excitations ($E_{\rm u}$) 
from 7  to 28 K (see Table 1).
Figure 1 (Top) shows, as an example, the 
imaging of the CH$_3$OH(2$_{\rm 1,2}$--1$_{\rm 1,1}$) A
line emission on
top of the CO (1--0) image (Gueth et al. 1996), which accurately outlines
the B0, B1, and B2 cavities opened by
the precessing jet driven by L1157-mm, which is
located at $\Delta$$\alpha$ = --25$\arcsec$ and
$\Delta$$\delta$ = +63$\farcs$5 with respect to the centre
of the Fig. 1 map. 
Figure A.1 shows the images of the whole of the CH$_3$OH 
line emission reported in Table 1. 
In addition, we detected and imaged eight acetaldehyde lines 
($E_{\rm u}$ = 14-23 K), which are also reported in Table 1.
Figure 1 (Bottom) shows the  
CH$_3$CHO(5$_{\rm 0,4}$--4$_{\rm 0,4}$) E map, while
Fig. A.2 is for all the other CH$_3$CHO images. 

Using the ASAI (Astrochemical Surveys At IRAM\footnote{www.oan.es/asai}; 
Lefloch et al. 2017, 2018) unbiased spectral survey 
obtained with the IRAM 30m antenna, we evaluated the missing flux
of the present NOEMA dataset. Comparison between the ASAI and NOEMA spectra
indicates that a large
fraction (about 60\%) of
the CH$_3$OH and CH$_3$CHO emission is filtered out, indicating that their emission is extended
across structures larger than $\sim$14$\arcsec$. 
Indeed, this matches the goal of the present project,
which is the analysis of the chemical content of the small structures associated
with multiple shocked regions, and without any contribution due to more extended emission.
In \S 4.3 and \S 5.1 the present results will be compared with those derived 
by the analysis based on single-dish observations, showing there is no bias
introduced by the missing flux.

\begin{figure}
\centerline{\includegraphics[angle=0,width=6.3cm]{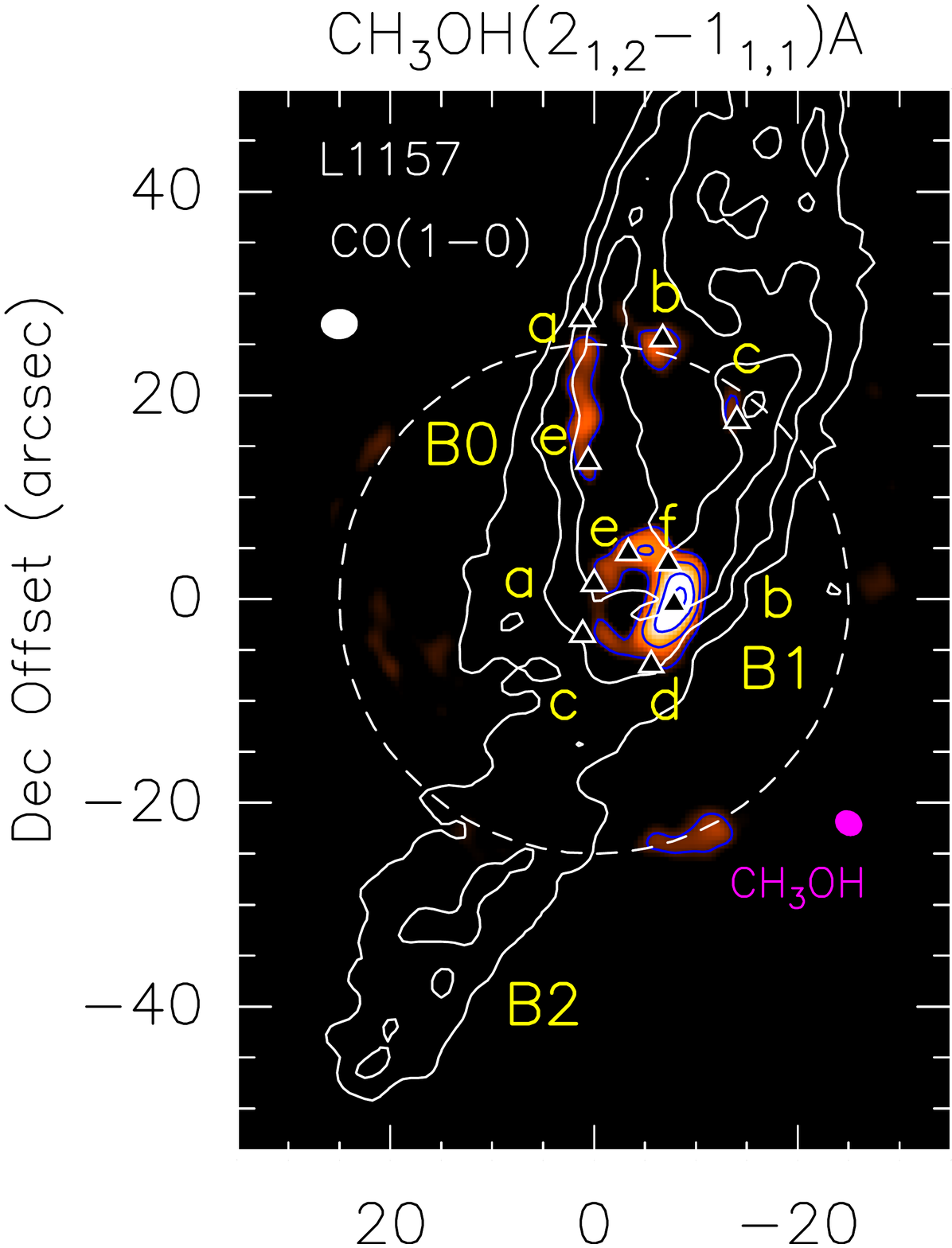}}
\vspace{0.2cm}
\centerline{\includegraphics[angle=0,width=6cm]{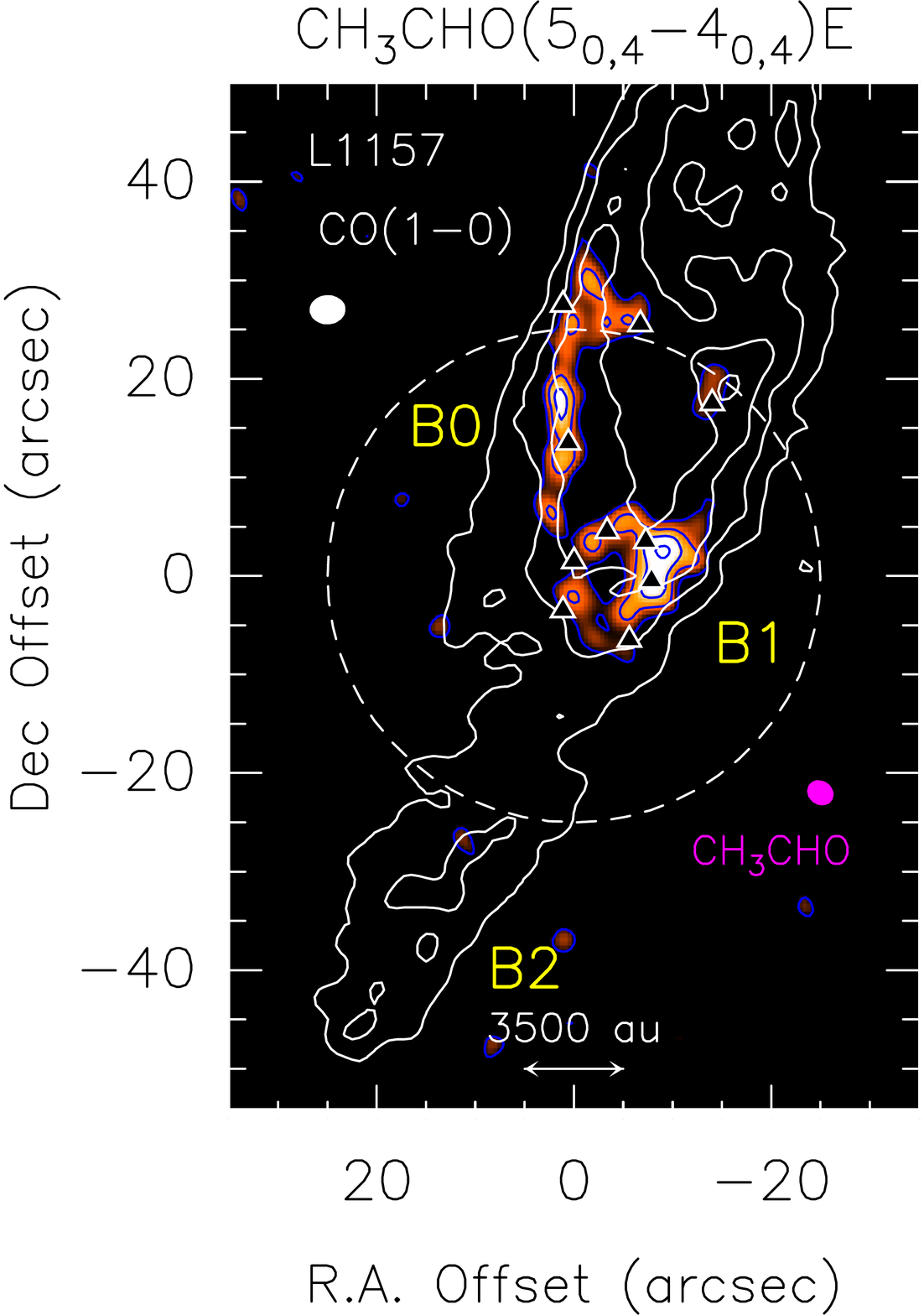}}
\caption{
The L1157 southern blueshifted lobe
in CO (1--0) (white contours; Gueth et al. 1996). 
The precessing jet
ejected by the central object L1157-mm (outside the frame and toward the northwest) excavated several clumpy cavities,
named B0, B1, and B2, respectively. The maps are centred at $\alpha({\rm J2000})$ = 20$^h$ 39$^m$ 10$\fs$2,
$\delta({\rm J2000})$ = +68$\degr$ 01$\arcmin$ 10$\farcs$5 ($\Delta$$\alpha$ = +25$\arcsec$ and
$\Delta$$\delta$ = --63$\farcs$5 from the L1157-mm protostar).
{\it Top panel:} Map (in colour scale) of the emission due to the  CH$_3$OH(2$_{\rm 1,2}$--1$_{\rm 1,1}$) A transition 
(integrated over the whole velocity range). The first contour and step are 3$\sigma$
(32 mJy beam$^{-1}$ km s$^{-1}$).
For the CO image, the first contour and step are
6$\sigma$ (1$\sigma$ = 0.5 Jy beam$^{-1}$ km s$^{-1}$) and 4$\sigma$, respectively. 
The dashed circle shows the primary beam of the CH$_3$OH image (64$\arcsec$). 
Yellow labels are for the B0 and B1 clumps (black triangles) previously identified using molecular tracers 
(see text, and e.g. Benedettini et al. 2007; Codella et al. 2009, 2017).
The white and magenta ellipses depict
the synthesised beams of 
the CO (3$\farcs$65 $\times$ 2$\farcs$96, PA=+88$\degr$), and CH$_3$OH (2$\farcs$97 $\times$ 2$\farcs$26, PA=--155$\degr$) 
observations, respectively.
{\it Bottom panel:} Same as top panel but for CH$_3$CHO (5$_{\rm 0,4}$--4$_{\rm 0,4}$) E. The first contour and step are 3$\sigma$
(7 mJy beam$^{-1}$ km s$^{-1}$). The synthesised beams of the  CH$_3$CHO image
are similar to those of the CH$_3$OH map.}
\label{maps}
\end{figure}

\begin{figure}
\centerline{\includegraphics[angle=90,width=9cm]{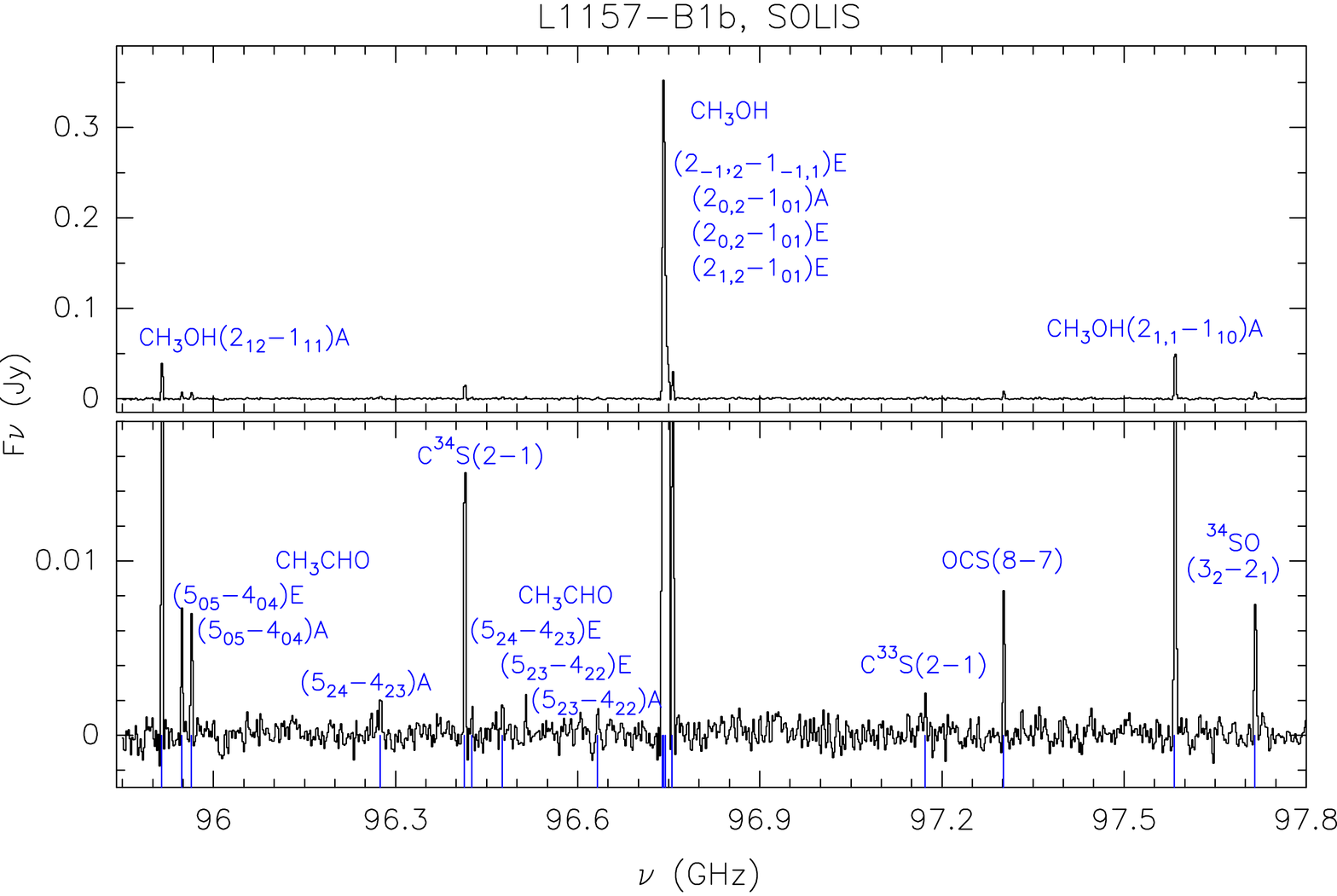}}
\centerline{\includegraphics[angle=90,width=8.9cm]{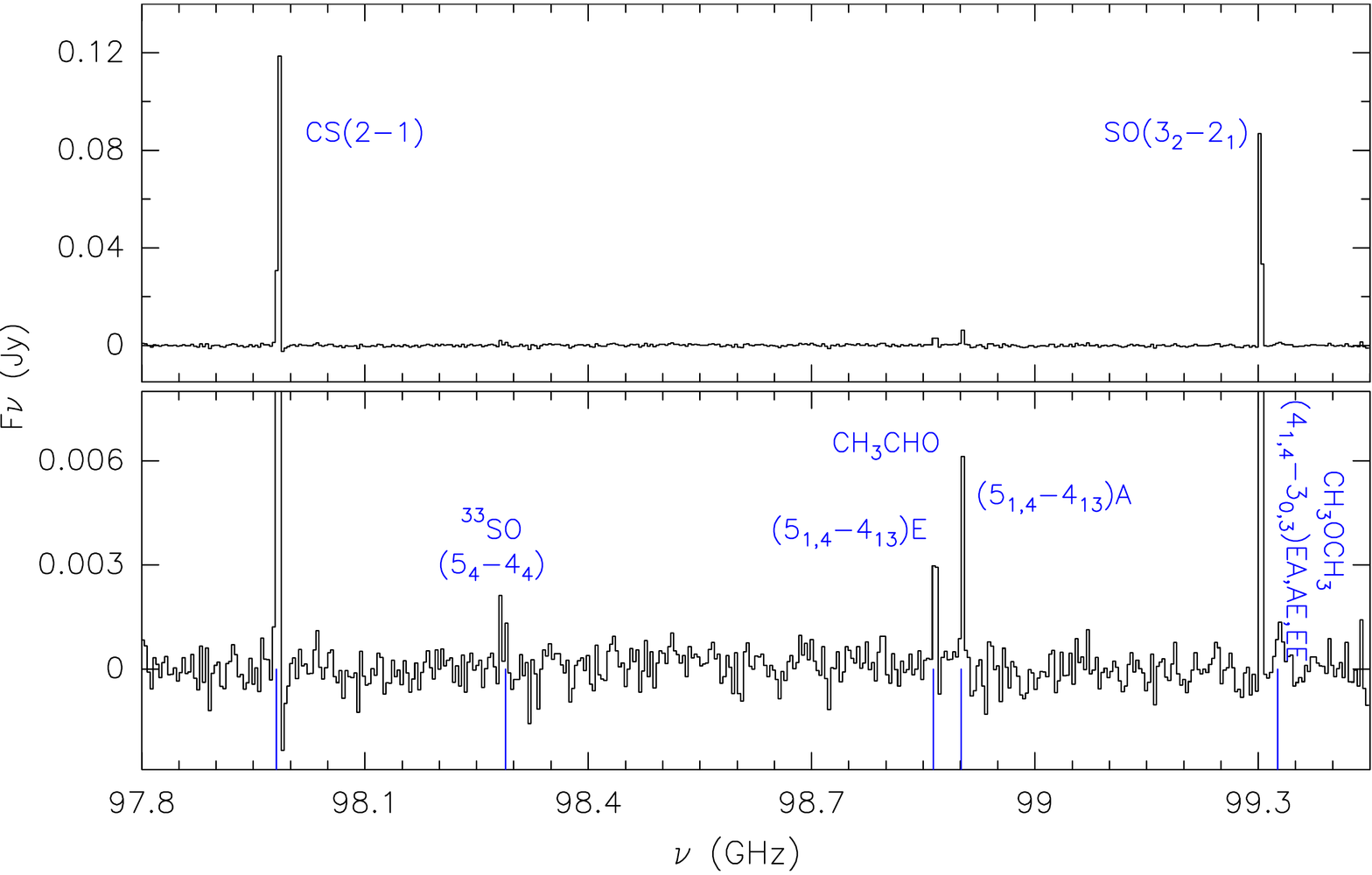}}
\caption{{\it Upper and Lower panels:} IRAM-WideX spectra (in $F_{\rm \nu}$ scale) extracted at
the L1157-B1b blueshifted position (see Fig. 1 and text).
The spectral resolution is 1.95 MHz (see Sect. 3).
A zoom-in in intensity is shown so that the weakest lines can be seen.
The transitions producing the emission lines are labelled in blue, with
the corresponding frequencies marked by vertical blue lines.
The detected CH$_3$OH and CH$_3$COH 
lines (analysed here) are reported in Table 1.}
\label{peakspectra}
\end{figure}

\subsection{CH$_3$OH and CH$_3$CHO spatial distribution}

The present NOEMA CH$_3$OH data improve the images and spectra 
based on IRAM-PdBI observations of the methanol spectral pattern at 97.94 GHz 
performed in 1997 using five antennas, and
with a synthesised beam of $\sim$ 5$\arcsec$,
as well as those of the 2$_{\rm 1,1}$--1$_{\rm 1,0}$ A line, with a spectral
resolution of 0.5 km s$^{-1}$ (Benedettini et al. 2007).  
The eight CH$_3$CHO images also allow us to perform a deeper analysis with respect to the
spatial distributions obtained by Codella et al. (2015), who stacked the 
CH$_3$CHO (7$_{0,7}$--6$_{0,6}$) E and A lines 
with an angular resolution of $\sim$ 2$\farcs$5. 

The present results are consistent with the previous observations that
show that the CH$_3$OH and CH$_3$CHO spatial distributions are
in good agreement; both trace the B0 eastern molecular cavity as well as
the northern arch-like structure containing B1a-e-f-b.  In particular,
the peak emission of both species is clearly associated with the B1b
clump.  In other words, although emission is observed also towards
B1c, the bulk of the CH$_3$OH and CH$_3$CHO emission does not come
from the southern B1 apex (Codella et al. 2017). Instead, they
preferentially trace the region associated with the youngest B1 shock
occurring next to the B1a position, where HDCO, a selective tracer of dust mantle release,
has been observed (Fontani et al. 2014),
as well as high-velocity SiO emission (Gueth et al. 1998; Podio et
al. 2016), locating the likely current sputtering of the dust
refractory cores due to the shock.  These findings are not surprising
for methanol, which is efficiently produced on the grain
surfaces by CO hydrogenation. On the other hand, they indicate that acetaldehyde is
either formed on the mantles as well or that it is quickly (in less than $\sim$ 1000 yr)
produced in the gas phase using simpler mantle products (as suggested
by Codella et al. 2015). This topic is further discussed in \S 5.2.

\begin{figure}
\centerline{\includegraphics[angle=0,width=8cm]{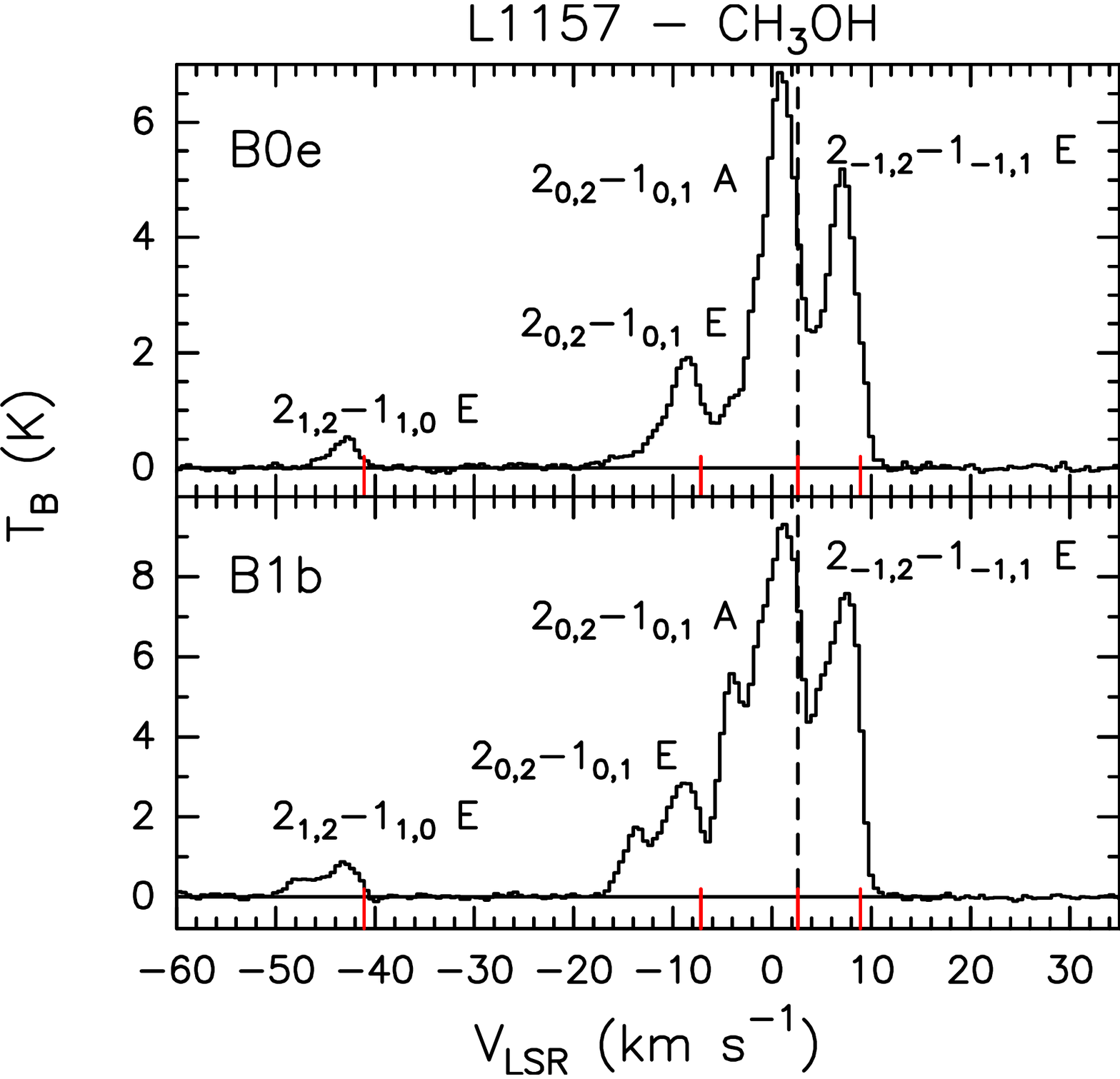}}
\vspace{-1cm}
\caption{CH$_3$OH emission spectra at 96.7 GHz (in $T_{\rm B}$ scale) extracted at
the B0e and B1b positions (see Fig. 1 and text).
The spectral resolution is 0.48 MHz, corresponding to 0.16 km s$^{-1}$.
The transitions producing the emission lines (see Table 1) are reported, with
the corresponding frequencies marked by small vertical red lines.
The spectra are centred at the frequency of the 2$_{\rm 0,2}$--1$_{\rm 0,1}$ A 
transition: 96741.38 MHz.
The dashed vertical line stands for the systemic velocity: +2.6 km s$^{-1}$
(e.g. Bachiller et al. 2001).}
\label{narrowspectra}
\end{figure}

\begin{figure}
\centerline{\includegraphics[angle=0,width=8cm]{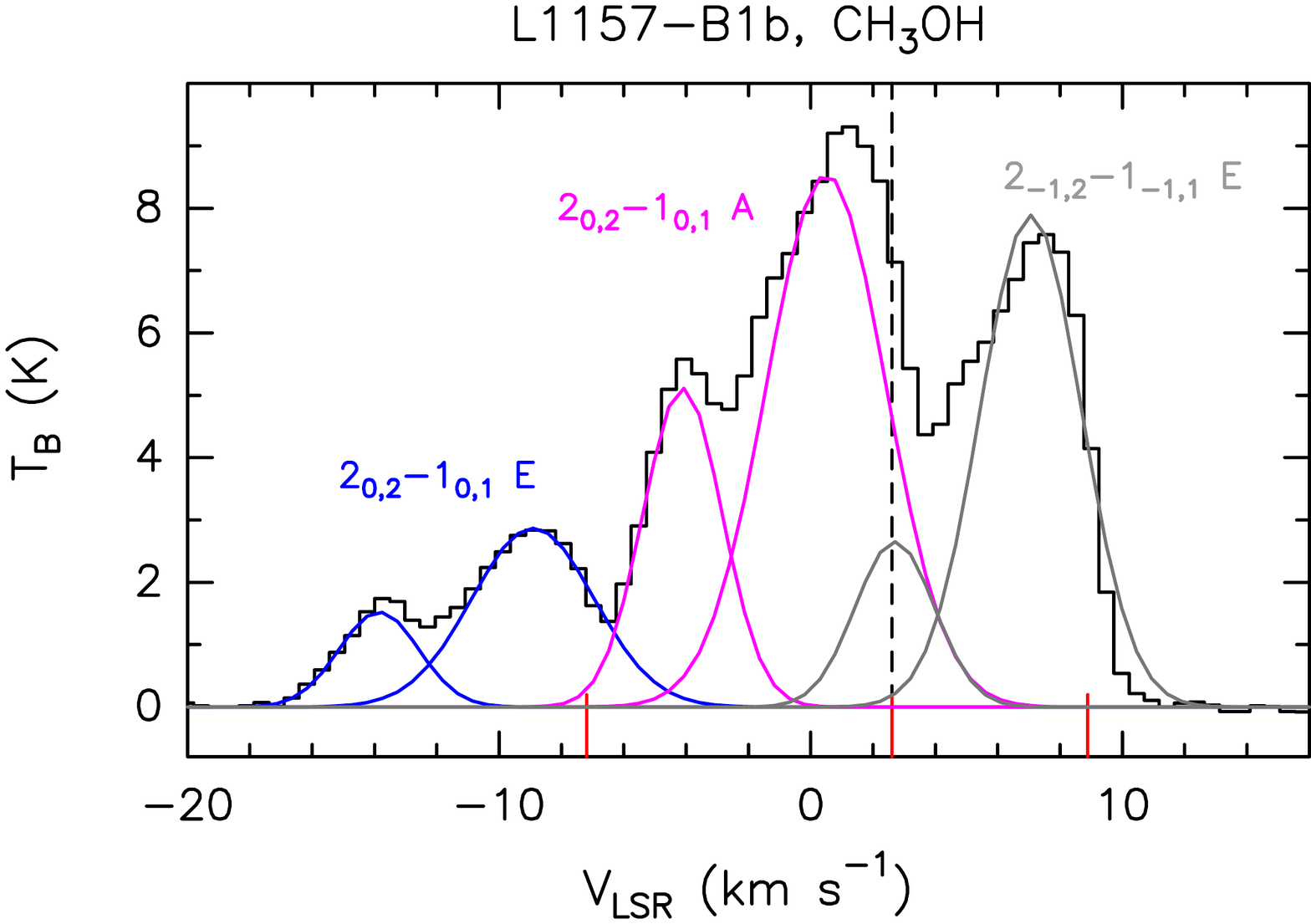}}
\vspace{-3cm}
\caption{Spectral fit of the 2$_{\rm -1,2}$--1$_{\rm -1,1}$ E (grey lines), 2$_{\rm 0,2}$--1$_{\rm 0,1}$ A (magenta), and
2$_{\rm 0,2}$--1$_{\rm 0,1}$ E (blue) triplets as observed towards B1b using the NOEMA narrow-band spectrometer.
The spectra are centred at the frequency of the 2$_{\rm 0,2}$--1$_{\rm 0,1}$ A 
transition: 96741.38 MHz. The transitions producing the emission lines (see Table 1) are reported, with
the corresponding frequencies marked by small vertical red lines. The dashed vertical line is for the systemic velocity: 
+2.6 km s$^{-1}$ (e.g. Bachiller et al. 2001). Each transition shows a main peak at  +0.5 km s$^{-1}$ plus
a secondary peak further blueshifted by $\sim$ 5 km s$^{-1}$.}
\label{multifit}
\end{figure}

\begin{table*}
\caption{CH$_3$OH and CH$_{3}$CHO emission lines observed using NOEMA and the WideX backend towards the L1157-B1b 
peak (see Fig. 2).}
\begin{tabular}{lcccccccc}
\hline
\multicolumn{1}{c}{Transition$^a$} &
\multicolumn{1}{c}{$\nu_{\rm 0}$ $^a$} &
\multicolumn{1}{c}{$E_{\rm u}$ $^a$} &
\multicolumn{1}{c}{$S\mu^2$ $^a$} &
\multicolumn{1}{c}{log(A$_{ij}$) $^a$} &
\multicolumn{1}{c}{rms} &
\multicolumn{1}{c}{$I_{int}$ $^b$} &
\multicolumn{1}{c}{$T_{\rm rot}$}&
\multicolumn{1}{c}{$N_{\rm tot}$}\\
\multicolumn{1}{c}{ } &
\multicolumn{1}{c}{(GHz)} &
\multicolumn{1}{c}{(K)} &
\multicolumn{1}{c}{(D$^2$)} &
\multicolumn{1}{c}{ } &
\multicolumn{1}{c}{(mK)}  &
\multicolumn{1}{c}{(K km s$^{-1}$)}  &
\multicolumn{1}{c}{(K)}  &
\multicolumn{1}{c}{(cm$^{-2}$)}  \\
\hline
\multicolumn{7}{c}{CH$_3$OH} & 10.0(1.1) & 1.3(0.3) $\times$ 10$^{16}$ \\
\hline
2$_{\rm 1,2}$--1$_{\rm 1,1}$ A & 95.91431 & 21 & 1.2  & --5.6  &  14 & 9.05(0.15) & & \\
2$_{\rm -1,2}$--1$_{\rm -1,1}$ E$^c$ & 96.73936 & 13 & 1.2 & --4.6  & 9 & 40.5(1.5)$^d$ & & \\ 
2$_{\rm 0,2}$--1$_{\rm 0,1}$ A$^c$ & 96.74138 & 7 & 1.6 & --5.6  & 9 &  57.2(1.5)$^d$ & & \\
2$_{\rm 0,2}$--1$_{\rm 0,1}$ E$^c$ & 96.74455 & 20 & 1.6 & --5.5  & 9 & 19.0(0.5)$^d$ & & \\
2$_{\rm 1,1}$--1$_{\rm 1,0}$ E & 96.75551 & 28 & 1.2 & --5.5  & 9 &  6.69(0.09) & & \\
2$_{\rm 1,1}$--1$_{\rm 1,0}$ A & 97.58280 & 22 & 1.2 & --5.6  & 11 &  12.82(0.13) & & \\
\hline
\multicolumn{7}{c}{CH$_{\rm 3}$CHO} & 8(1) & 7(3)  $\times$ 10$^{13}$ \\
\hline
5$_{\rm 0,5}$--4$_{\rm 0,4}$ E & 95.94744 & 14 &  63.2 & --4.5  & 5 &  1.71(0.14) & & \\ 
5$_{\rm 0,5}$--4$_{\rm 0,4}$ A & 95.96346 & 14 & 63.2 & --4.5  & 7 &  1.65(0.15) & & \\ 
5$_{\rm 2,4}$--4$_{\rm 2,3}$ A & 96.27425 & 23 & 53.1 & --4.6  & 8 &  0.65(0.15) & & \\ 
5$_{\rm 2,4}$--4$_{\rm 2,3}$ E & 96.42561 & 23 & 52.8 & --4.6  & 9 &  0.40(0.08) & & \\  
5$_{\rm 2,3}$--4$_{\rm 2,2}$ E & 96.47552 & 23 & 52.8 & --4.6  & 11 &  0.44(0.16) & & \\
5$_{\rm 2,3}$--4$_{\rm 2,2}$ A & 96.63266 & 23 & 53.1 & --4.6  & 12 &  0.42(0.09) & & \\
5$_{\rm 1,4}$--4$_{\rm 1,3}$ E & 98.86331 & 17 & 60.7 & --4.5  & 12 &  1.42(0.19) & & \\       
5$_{\rm 1,4}$--4$_{\rm 1,3}$ A & 98.90094 & 17 & 60.7 & --4.5  &  9 &  1.50(0.12) & & \\
\hline
\end{tabular}

$^a$ Frequencies and spectroscopic parameters were extracted from the Jet Propulsion Laboratory molecular 
database (Pickett et al. 1998) for CH$_3$CHO and CH$_3$OCHO. 
Upper level energies refer to the ground state of each symmetry. 
$^b$ Integrated over the whole velocity emission range from the WideX spectra extracted at the CH$_{\rm 3}$CHO
and CH$_3$OH emission peak (B1b; see text). In case of non-detections we report the 3$\sigma$ limit. 
We note that the errors on the integrated areas $I_{int}$ are propagated using the r.m.s. and the spectral resolution of the spectra. 
In addition, the uncertainty on the absolute flux scale due to calibration is $\leq$ 15\% (see Sect. 3).
$^c$ The three lines at 96.74 GHz have been deblended using the 0.5 km s$^{-1}$ spectral resolution (see Sect. 3.2 and Fig. 4). \\
\end{table*}

\subsection{Line spectra}

Figure 2 shows the spectra
extracted from the CH$_3$OH and  CH$_3$CHO emission peaks at  the B1b position.
In addition to methanol and acetaldehyde, in B1b we observe emission at 99.3 GHz due to another
iCOM, dimethyl ether (CH$_3$OCH$_3$). 
In particular, the detected line is due to the 4$_{\rm 1,4}$--3$_{\rm 0,3}$ transitions of the
EA, AE, EE, and AA isomers, blended at the $\sim$ 2 MHz spectral resolution (see Fig. 2). 
Finally, we detect several S-bearing species (CS, SO, and OCS isotopologues),
which will be analysed in a forthcoming paper on sulphur chemistry (Feng et al., in preparation).  

\begin{figure}
\centerline{\includegraphics[angle=0,width=8cm]{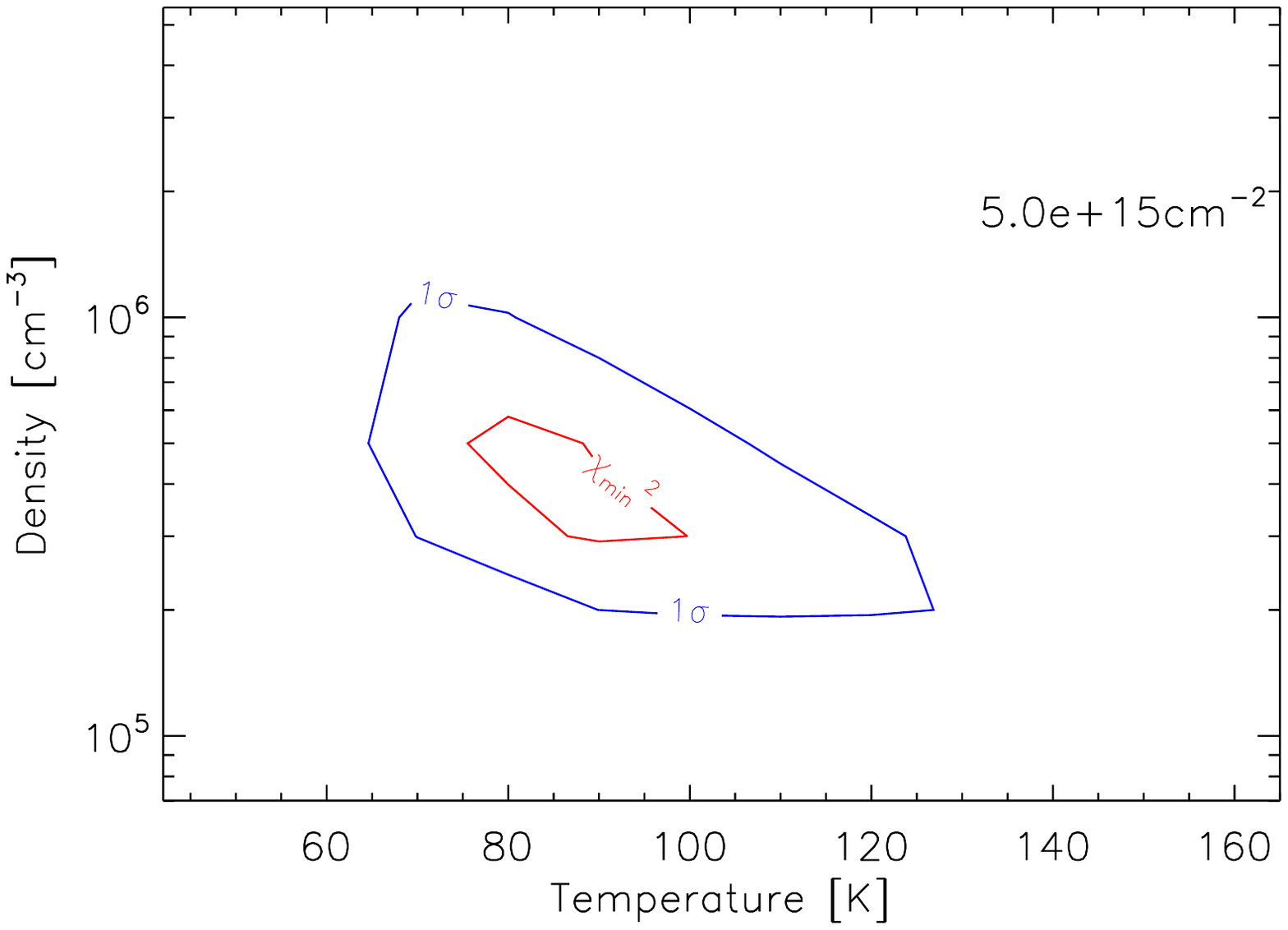}} 
\vspace{-6cm}
\centerline{\includegraphics[angle=0,width=8cm]{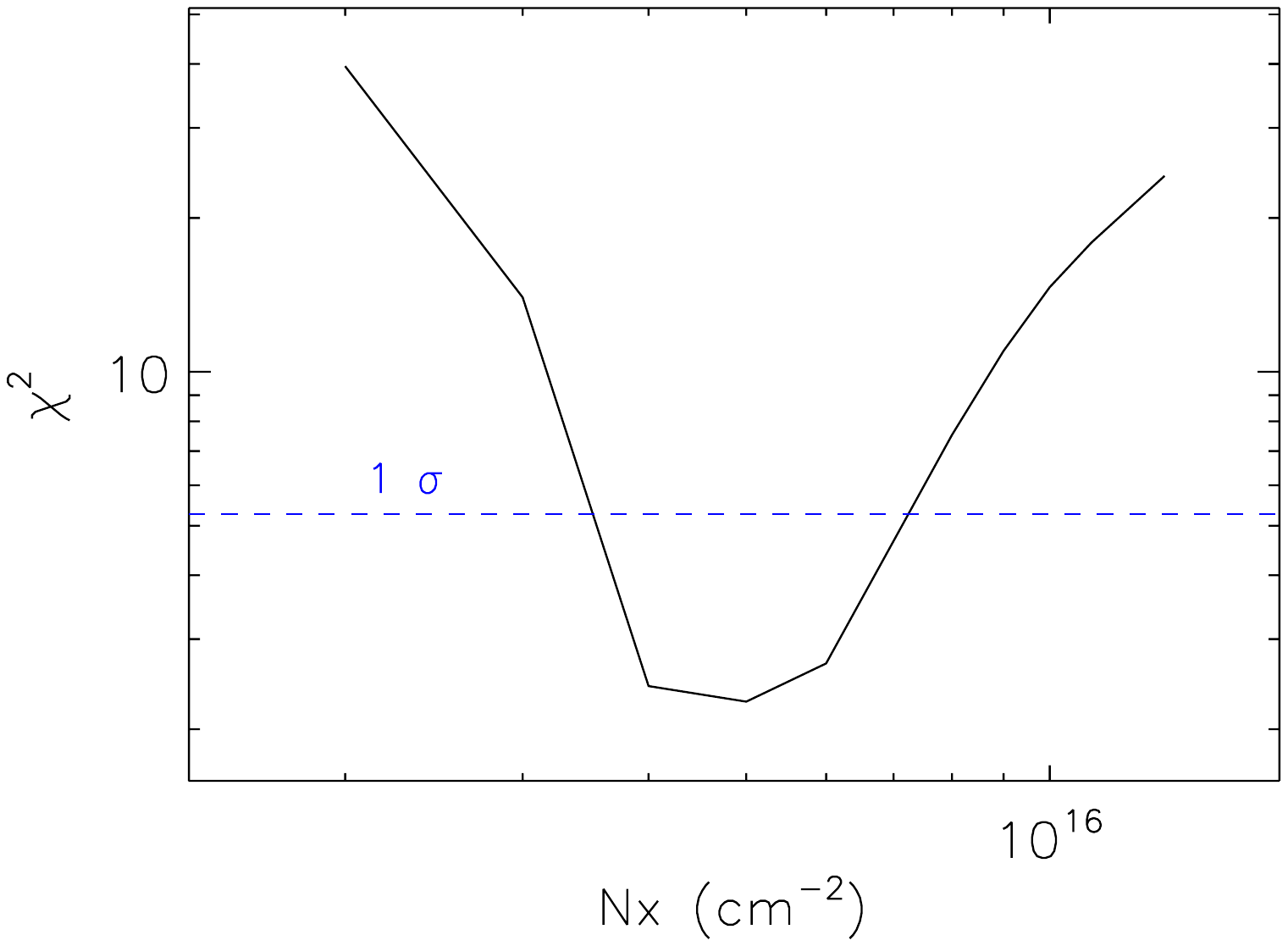}}
\vspace{-6cm}
\centerline{\includegraphics[angle=0,width=8cm]{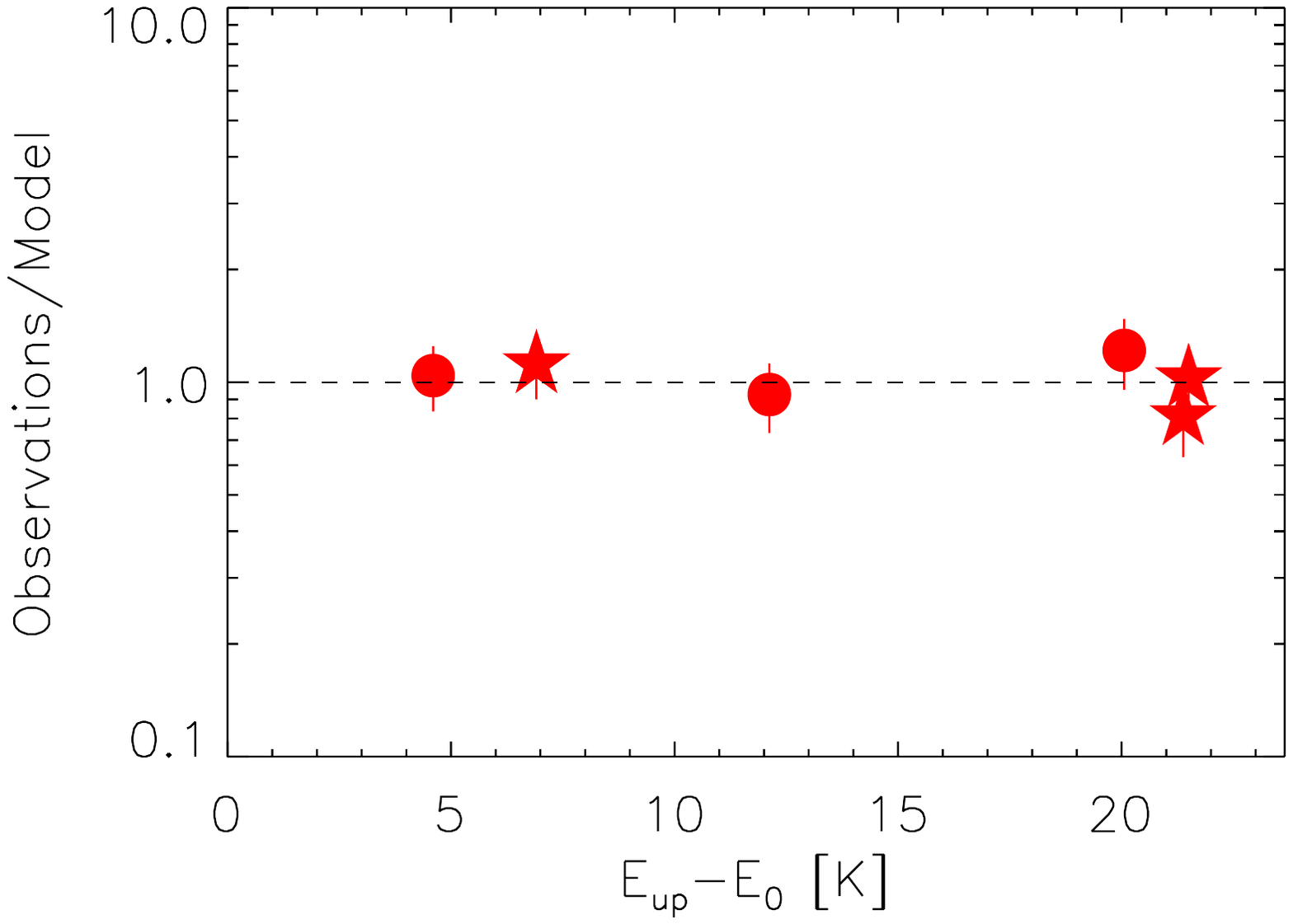}} 
\vspace{-5cm}
\caption{{\it Upper panel:} Density-temperature contour plot of
  $\chi^{2}$ obtained considering the non-LTE LVG-model-predicted and
  observed intensity of all the A and E CH$_3$OH emission lines
  detected (beam averaged) towards the B1b clump.  The best fit is
  obtained with N(CH$_3$OH) = 5 $\times$ 10$^{15}$ cm$^{-2}$,
  $T_{\rm kin}$ = 90 K, and $n_{\rm H_{2}}$ = $4\times$ 10$^{5}$
  cm$^{-3}$. {\it Middle panel}: $\chi^{2}$--N(CH$_3$OH) distribution,
  identifying the minimum $\chi^{2}$ value (2.4). 
  {\it Lower panel:} Ratio between observations and model
  predictions of the CH$_3$OH A (circles) and E (stars) line intensities
  as a function of the upper level energy of the lines.}
\label{lvg}
\end{figure}

For each detected line,  in addition to the spectroscopic parameters, Table 1 reports the intensity integrated 
over the whole velocity emission range (down to velocities blueshifted by 15 km s$^{-1}$). 
The 2 MHz spectral resolution of the WideX backend does not allow us to 
investigate the iCOMs kinematics. However, we can confirm that all the spectra 
are blueshifted, in agreement with all the molecular lines so far observed towards L1157-B1 
(e.g. Bachiller et al. 2001; Lefloch et al. 2017). 
In addition, the CH$_3$OH spectral pattern at 96.7 GHz (see Table 1) 
has been observed with a 0.48 MHz (0.16 km s$^{-1}$) spectral resolution. Figure 3 reports the
spectra as observed towards the B0e and B1b clumps: four transitions of the 
2$_{\rm k,k}$--1$_{\rm k,k}$ pattern are reported, with
the corresponding frequencies marked by small vertical red lines.
The spectra are centred at 96741.38 MHz, which is the frequency of the 
2$_{\rm 0,2}$--1$_{\rm 0,1}$ A line.
The methanol lines extracted from the B0e position have a peak velocity of  $\sim$ +0.5 km s$^{-1}$
($V_{\rm LSR}$ = +2.6 km s$^{-1}$), 
and are about 3--4 km s$^{-1}$ in width (FWHM), with
evidence of bluer wings.
The B1b spectrum is definitely more complex showing in addition a secondary peak
at about --4 km s$^{-1}$. This high-velocity peak was tentatively detected by Benedettini et al. (2013) using
the CH$_3$OH(3$_{\rm k,k}$--2$_{\rm k,k}$) emission at 2mm. In the present dataset, a 
clear Gaussian-like secondary peak is emerging, definitely offset in velocity from the mean peak. 
These findings confirm that the northern part of the B1 structure is associated
with young shocked gas, here traced by methanol molecules injected into the gas
phase at high velocities. 

Figure 4 shows the spectral fit of the 2$_{\rm -1,2}$--1$_{\rm -1,1}$ E, 2$_{\rm 0,2}$--1$_{\rm 0,1}$ A, and
2$_{\rm 0,2}$--1$_{\rm 0,1}$ E  triplets as observed towards B1b. The velocity scale is based on the rest frequency of
the brighter line, 2$_{\rm 0,2}$--1$_{\rm 0,1}$ A, at 96741.38 MHz.
In order to perform the fit, we assumed that the three low-excitation methanol transitions 
are associated with the same spectral profile. As a consequence, the rest frequencies of the transitions give us
the expected velocity shifts among the three lines (see the red small vertical segments in Fig. 4).
In addition, both 2$_{\rm 0,2}$--1$_{\rm 0,1}$ E and A lines (blue and magenta) show a secondary peak (3.0 km s$^{-1}$ broad)
offset by 5 km s$^{-1}$ with respect the main one (which has a line width of 4.6 km s$^{-1}$).
In conclusion, we fixed the position and the line width of the secondary peak expected for the 
2$_{\rm -1,2}$--1$_{\rm -1,1}$ E line (grey). The overall fit is satisfactory, with a residual with an rms of 480 mK.

\begin{figure}
\centerline{\includegraphics[angle=0,width=8cm]{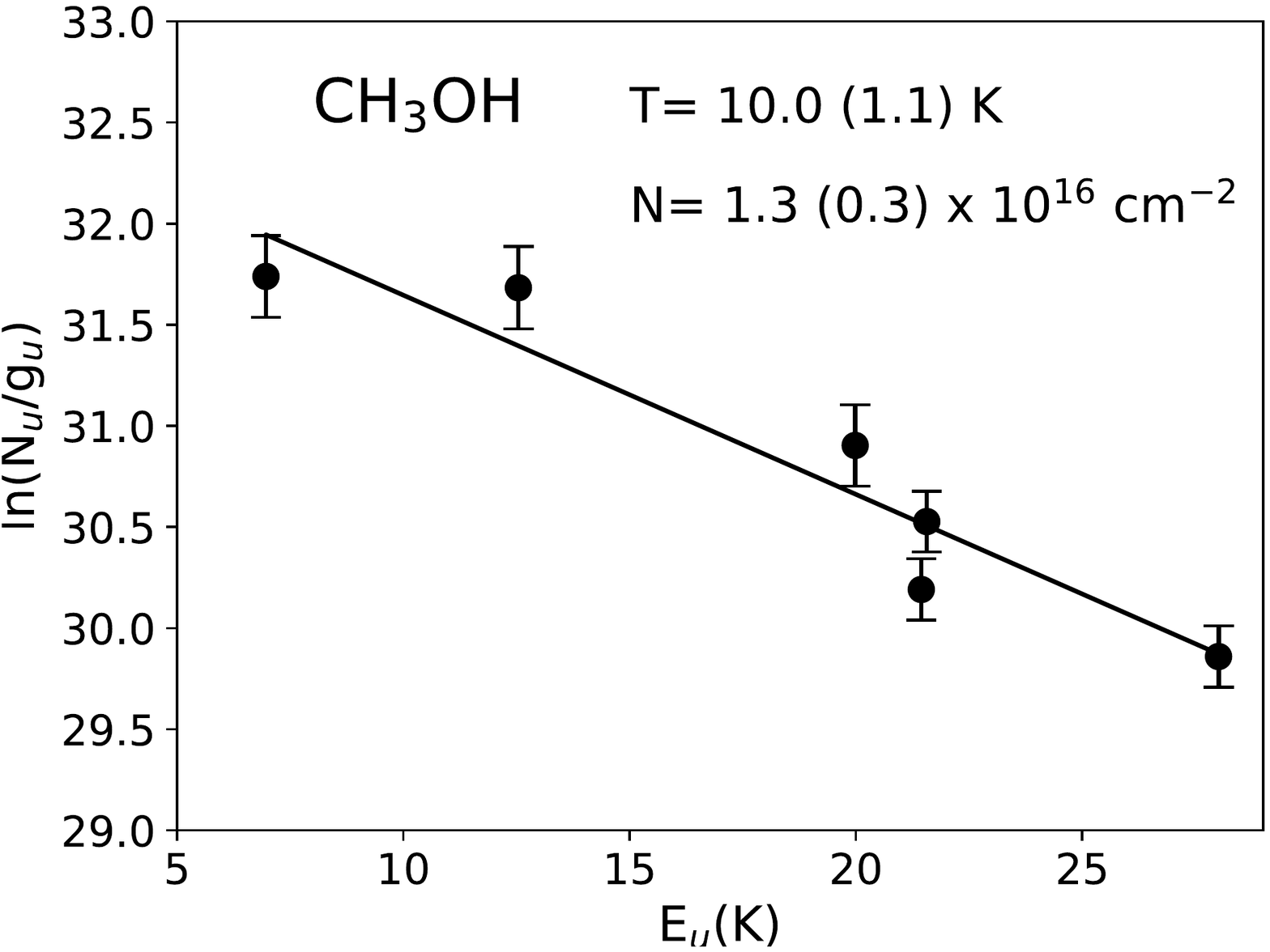}} 
\vspace{-5cm}
\centerline{\includegraphics[angle=0,width=8cm]{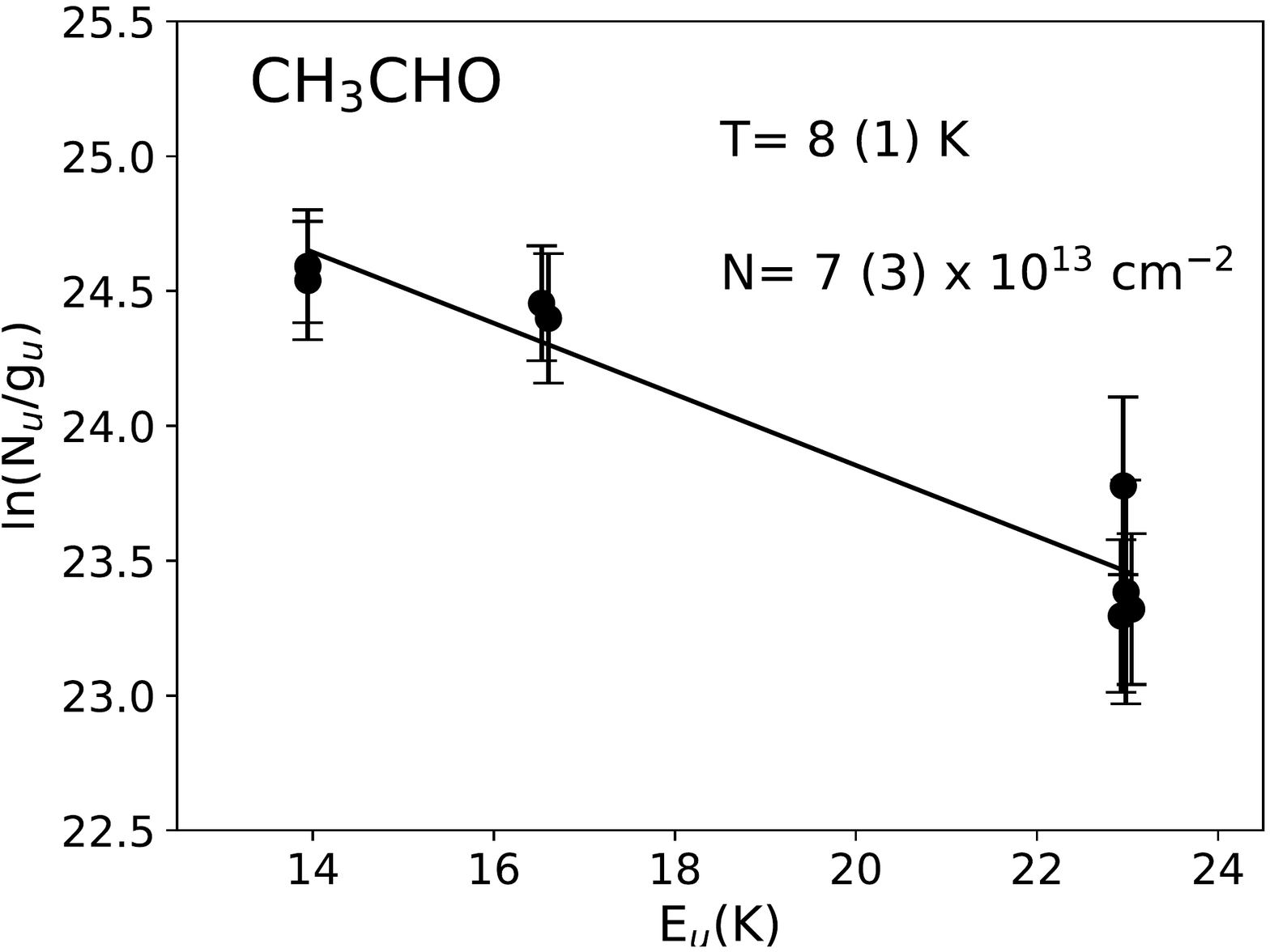}} 
\vspace{-2cm}
\caption{Rotational diagrams for CH$_3$OH ({\it upper panel}), and
  CH$_3$CHO ({\it lower panel}) derived using the emission lines
  observed towards B1b (see Table 1 and Fig. 3).  The parameters
  N$_{u}$, g$_{u}$, and $E_{\rm up}$ are the column
  density, the degeneracy, and the energy (with respect to the ground
  state of each symmetry), respectively, of the upper level. The derived values of
  the rotational temperature are reported in the panels.}
\label{RD}
\end{figure}

\subsection{Large Velocity Gradient analysis and rotational diagrams: column densities and abundances}

Given the lack of CH$_{3}$CHO collisional rates in the literature, we
only analysed the CH$_3$OH emission with the non-local
thermodinamic equilibrium  (NLTE) large velocity gradient (LVG) approach
using the model described in Ceccarelli et al. (2003).  We used the
CH$_{3}$OH-H$_{2}$ collisional coefficients computed by Rabli \&
Flower (2010) and provided by the BASECOL database (Dubernet et
al. 2013). In the computations, we assumed a H$_2$ ortho-to-para ratio
of one (Nisini et al. 2010).  We run grids of models varying
the kinetic temperature ($T_{\rm kin}$) from 50 to 300 K, the volume
density ($n_{\rm H_2}$) from 10$^{4}$ to 10$^{8}$ cm$^{-3}$, and the
methanol column density (N(CH$_3$OH)) from 10$^{14}$ to 10$^{18}$
cm$^{-2}$.  The best fit is obtained by the minimum $\chi^{2}$ in the
three parameters. The errors on the observed line intensities
  have been obtained by propagating the spectral rms  with the
  uncertainties due to calibration (15\%).

Figure 5 shows: (i) the $\chi^{2}$ contour plot in the
  $T_{\rm kin}$--$n_{\rm H_2}$ plane obtained with the best fit of the
  methanol column density N(CH$_{3}$OH) = $5 \times$ 10$^{15}$
  cm$^{-2}$ (3--8 $\times$ 10$^{15}$ cm$^{-2}$, including
  uncertainties), (ii) the $\chi^{2}$--N(CH$_{3}$OH) distribution,
  where it is possible to identify the minimum $\chi^{2}$ value (2.4),
  and (iii) the ratio between the measured and predicted intensity of
  the observed lines. Overall, the best fit is very good, with a
  reduced $\chi^{2}$ equal to 1.2, which for two degrees of freedom
  corresponds to a probability of 30\% to exceeding $\chi^{2}$
(i.e. approximately the standard 1 $\sigma$). 
The methanol line opacities
are predicted to be less than 0.1. The volume density is well
constrained, $n_{\rm H_2}$ = $4 \times$ 10$^{5}$ cm$^{-3}$ (2--10
$\times$ 10$^{5}$ cm$^{-3}$), in agreement with the high densities
($\geq$ 10$^{5}$ cm$^{-3}$) inferred toward the molecular cavities by
G\'omez-Ruiz et al. (2015) from a CS multiline analysis.  The kinetic
temperature is also well constrained, $T_{\rm kin}$ = 90 K (70--130
K), again in agreement with the temperatures of the B1 cavity as
derived by Lefloch et al. (2012) using a CO multiline single-dish
analysis. We note that a recent LVG analysis of methanol emission as
  observed using single-dish towards a large number of protostellar
  jet-driven shocks (Holdship et al. 2019) shows volume densities in
  the 10$^{5}$--10$^{6}$ cm$^{-3}$ range and kinetic temperatures
  between 20  and 60 K,  values consistent with what is found in
  L1157-B1.

The analysis  described immediately above allows us to constrain the density
and temperature of the methanol-emitting gas. However, in order to
derive the relative abundances of methanol and acetaldehyde, we
carried out a rotational diagram (RD) analysis for both molecules,
where a LTE population and optically thin lines are assumed (in
agreement with the NLTE LVG analysis of methanol described above).
Under these assumptions, for a given molecule, the relative population
distribution of all the energy levels is described by a Boltzmann
temperature, that is the rotational temperature $T_{\rm rot}$.
As demonstrated by  Lefloch et al. (2102), for example, by analysing CO emission from extended
outflowing gas in L1157, the column density obtained with RD gives results that are in relatively good agreement with  a NLTE LVG analysis, usually within a factor two, 
as is also the case for our analysis.

Figure 6 shows the RD of CH$_3$OH, which provides a column density of
(1.3$\pm$0.3) $\times 10^{16}$ cm$^{-2}$, close to what was found using
the LVG method. The rotational temperature is (10.0$\pm$1.1) K, in
agreement with what was found using previous interferometric and single
dish measurements of CH$_3$OH lines in the same excitation range (12
K; Bachiller et al. 1995; Benedettini et al. 2007; Codella et
al. 2010).  Thanks to the many CH$_3$CHO detected lines (covering an
upper level energy $E_{\rm u}$ range similar to that traced by the
methanol lines) we can carry out the RD analysis also for
acetaldehyde, obtaining, as for CH$_3$OH, a low rotational
temperature, (8$\pm$1) K, and a total column density N(CH$_{3}$CHO) =
(7$\pm$3) $\times 10^{13}$ cm$^{-2}$.  Lefloch et al. (2017) report
for CH$_{3}$CHO a rotational temperature of 17 K, but this was measured
including transitions with $E_{\rm u}$ up to 94 K.  Figure 7 shows the
spatial distribution of the CH$_3$CHO (upper panel) and CH$_3$OH
(lower panel) rotational temperature in the B0 and B1 regions of the
L1157 outflow. We note that the CH$_3$CHO rotational temperature map covers
a smaller portion of the L1157 region due to the weakness of the
acetaldehyde lines with $E_{\rm u}$ = 23 K.  Considering the
uncertainties (B0e and B1e: $\sim$ 6 K; B1b: $\sim$ 3 K) there is no
significant difference between the rotational temperature of the two
species or spatial trends throughout the L1157-B1 structure.

\begin{figure}
\centerline{\includegraphics[angle=0,width=8cm]{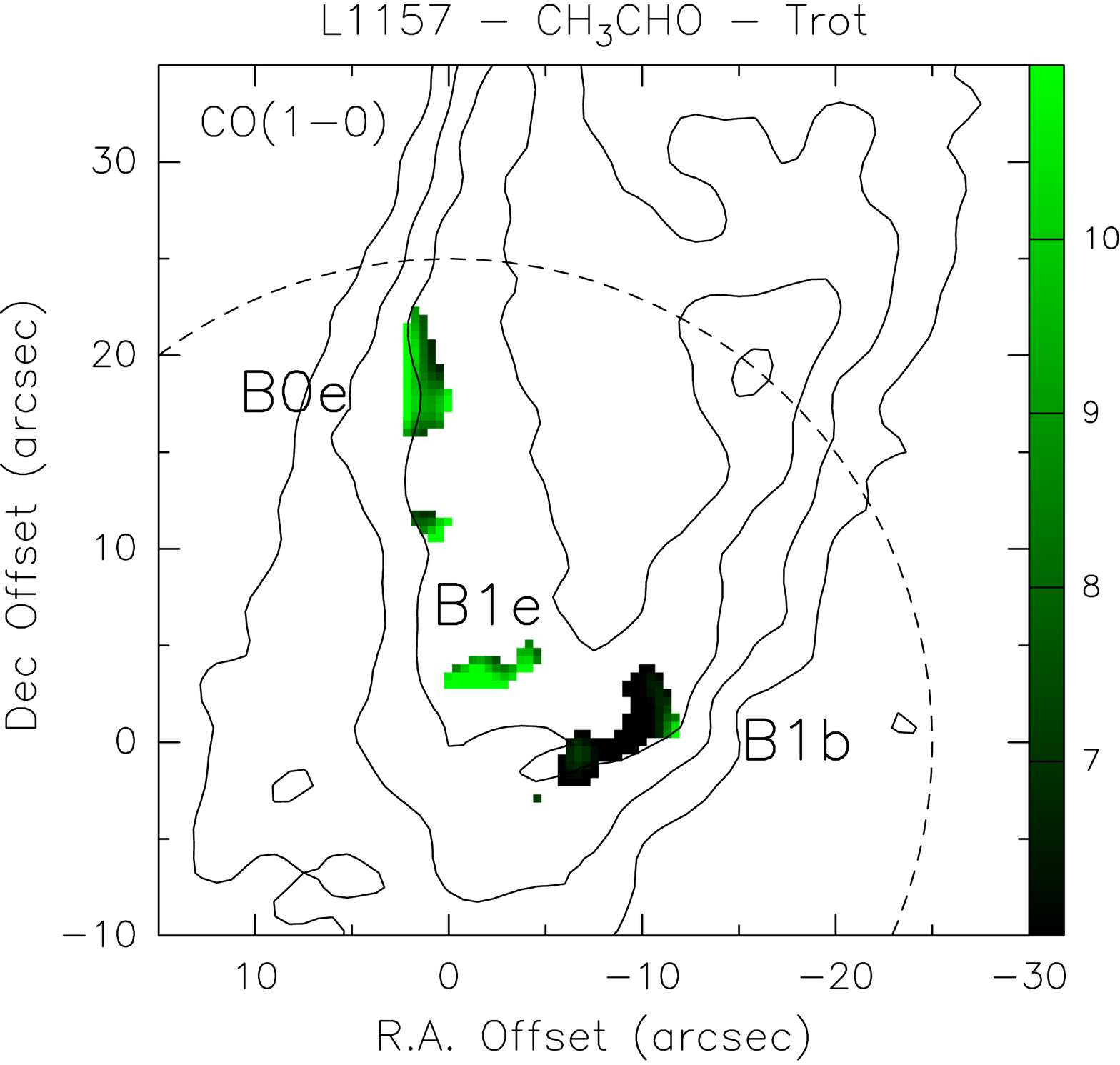}}
\vspace{-3cm}
\centerline{\includegraphics[angle=0,width=8cm]{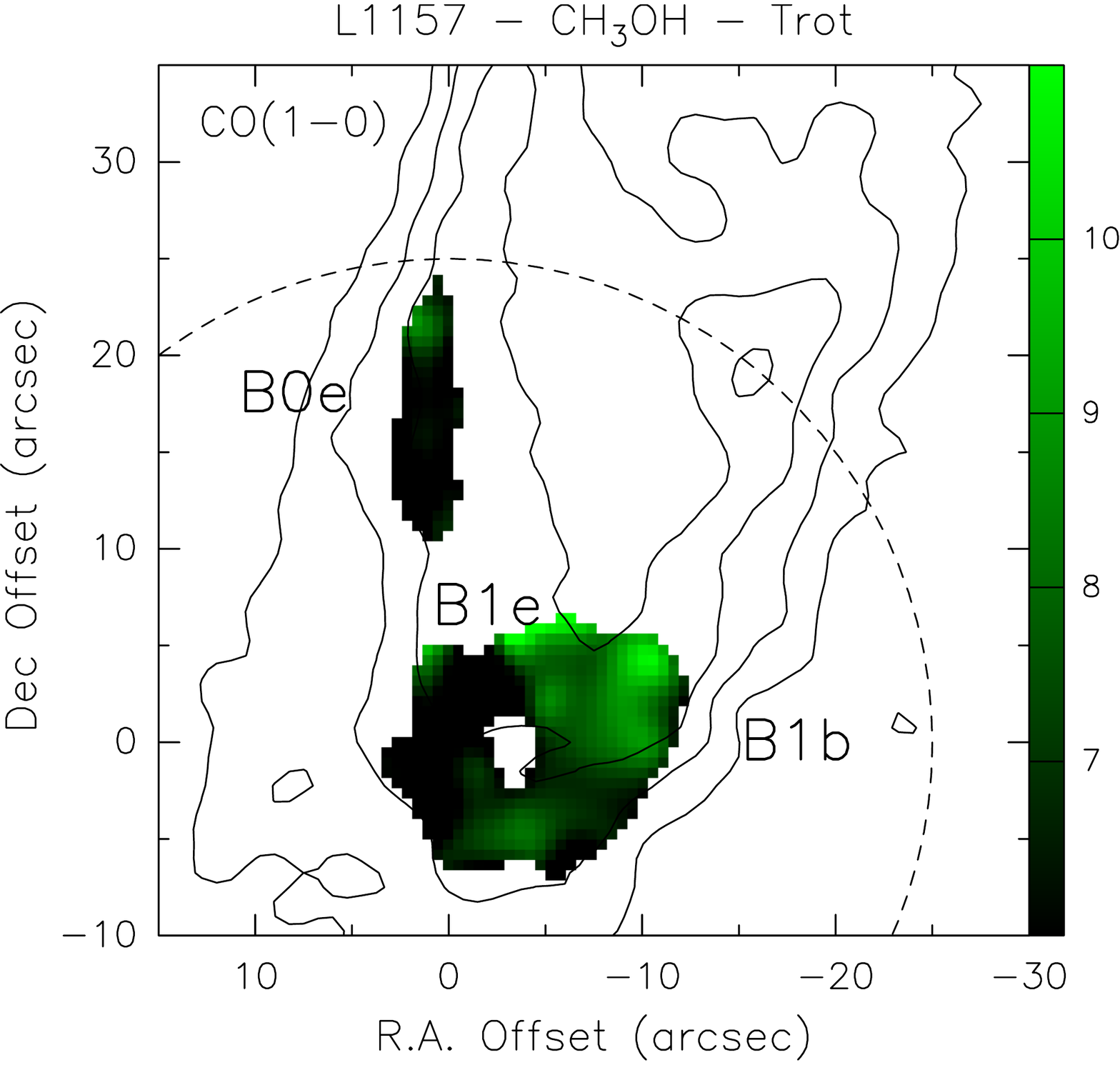}}
\vspace{-2cm}
\caption{ CH$_3$CHO ({\it upper panel}) and CH$_3$OH ({\it lower
    panel}) rotational temperature ($T_{\rm rot}$) map (in K scale) of
  the B0 and B1 region of the L1157 outflow (derived from the NOEMA
  images, where the emission is at least 3$\sigma$), overlaid on the
  CO (black contours; Gueth et al. 1996) emission map. Uncertainties
  on $T_{\rm rot}$ are $\sim$ 3 K towards B1b, and $\sim$ 6 K for B1e
  and B0. Symbols are as in Fig. 1.}
\label{Tex}
\end{figure}

\begin{figure}
\centerline{\includegraphics[angle=0,width=8cm]{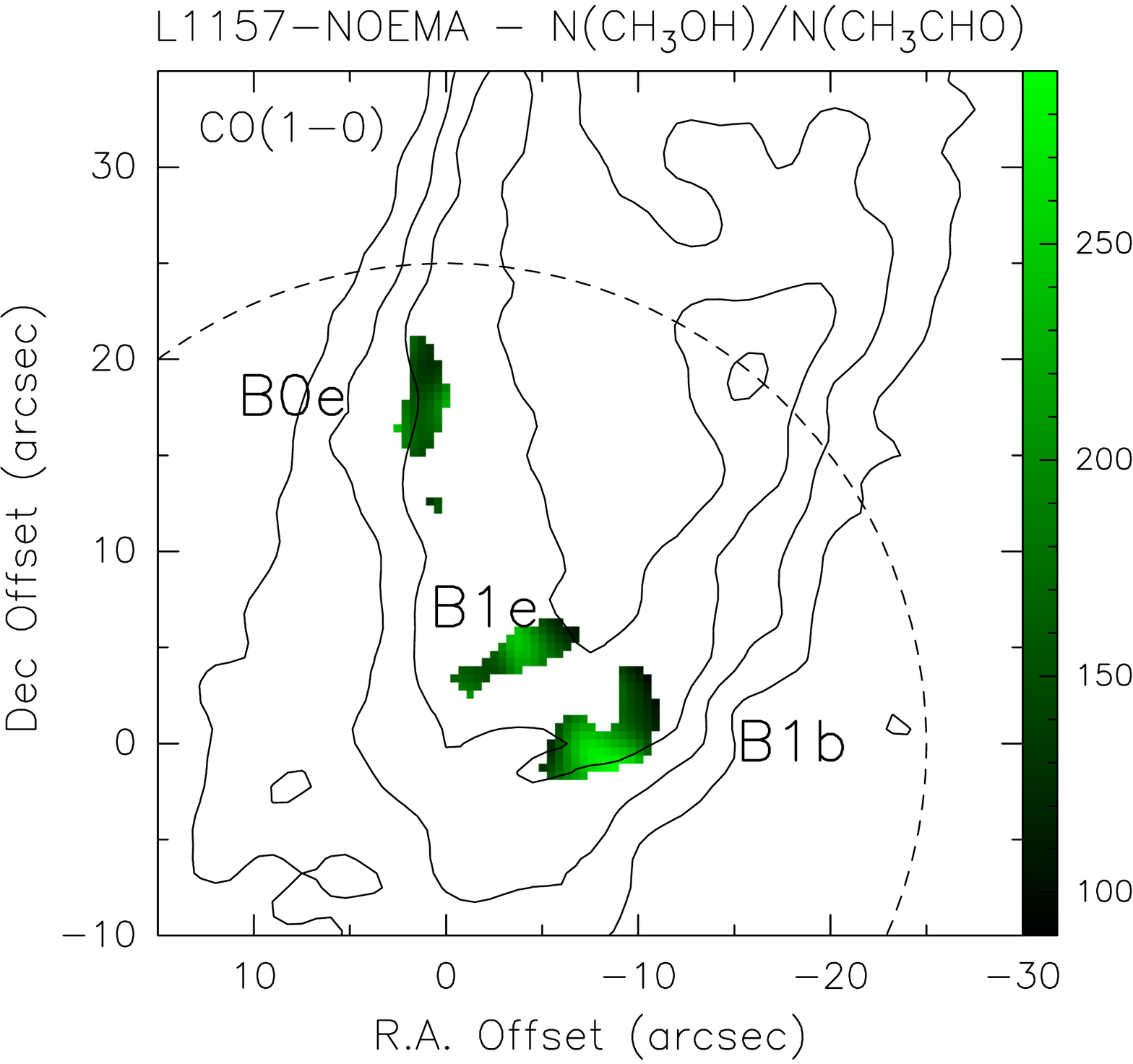}}
\vspace{-2cm}
\caption{Map of the ratio between the CH$_3$OH and CH$_3$CHO column
  density of the B0 and B1 region of the L1157 outflow (derived from
  the NOEMA images, where the emission is at least 3$\sigma$),
  overlaid on the CO (black contours; Gueth et al. 1996) emission
  map. Uncertainties are $\simeq$ 30\% (i.e. 30--100 in absolute values). Symbols are as in
  Fig. 1.}
\label{Tex}
\end{figure}

Finally, we computed the abundances of the various species using the
values of their respective column densities divided by the H$_2$
column density, which was derived from the CO column density in the B1 molecular
cavity (N$_{CO}$ = 2 $\times$ 10$^{17}$ cm$^{-2}$: Lefloch et
al. 2012) assuming the standard [CO]/[H] = 5 $\times$ 10$^{-5}$
value.  In this way, we obtain methanol and acetaldehyde abundances
with respect to H equal to 6.5 $\times$ 10$^{-6}$ (1.2--12 $\times$ 10$^{-6}$,
considered the uncertainties) and 3.5 $\times$ 10$^{-8}$ (2--5
$\times$ 10$^{-6}$), respectively, in good agreement with that derived
by CC17 for acetaldehyde (1--3 $\times$ 10$^{-8}$) using a coarser
angular resolution (5$\arcsec$--6$\arcsec$).

\section{Discussion}\label{sec:discussion}

\subsection{The CH$_3$OH to CH$_3$CHO abundance ratio}

Codella et al. 2017 showed that the spatial distribution of the measured
abundance ratio of iCOMs can lead to very stringent constraints on how
they are formed and destroyed.  In particular, these latter authors showed that the
formamide to acetaldehyde abundance ratio towards the L1157-B1 shock
demonstrates a gas-phase formation route of formamide (e.g. Skouteris
et al. 2017). Further analysis of the abundance
ratio of acetaldehyde with respect to methanol is therefore worthwhile, which can be safely
assumed to form on dust mantles and then injected into the gas
phase. Based on the interferometric maps showing that both species are
populating the same gas portion, the ratio between their column
densities can be used to derive their abundance ratio.  The ratio of CH$_3$OH
to CH$_3$CHO  is about 1.5 $\times$ 10$^{2}$ and 1.9 $\times$
10$^{2}$ as measured towards the B0e and B1b peaks, respectively, and
it varies by less than a factor of three throughout the whole B0--B1
structure, as shown in Fig. 8.  Considering that the uncertainty on these measurements is
about 30\%, there is no evidence for significant
variation of this latter ratio in the observed region.  Furthermore, these
measurements are consistent with those derived using the ASAI IRAM
30m single-dish survey (1--3 $\times$ 10$^{2}$; Codella et al. 2015,
Lefloch et al. 2017). We note that Holdship et al. (2019) measured
CH$_3$OH/CH$_3$CHO abundance ratios towards a sample of shocks using
the IRAM 30m antenna, finding higher values  (by a factor of six on average): 0.4--1.3 $\times$
10$^{3}$; high-spatial-resolution surveys are needed to confirm these
values.

It is interesting to check whether the CH$_3$OH/CH$_3$CHO abundance
ratios derived for L1157-B0 and L1157-B1 are different with respect to
measurements towards known Class 0 hot corinos, which rather trace
the inner 100 au from the protostar where the temperature is high
enough (at least 100 K) to thermally evaporate the dust-mantle
products into the gas phase, or where grain mantles are sputtered
by shocks (see below).  If we only include in the comparison those
sources imaged with high-spatial-resolution interferometric observations (to be
sure of a proper comparison between the CH$_3$OH and CH$_3$CHO spatial
distributions), the sample size is limited (see Table 2): HH212-mm (Lee et
al. 2017, 2019, Bianchi et al. 2017, Codella et al. 2018),
IRAS16293-2422B (J\o{}rgensen et al. 2016, 2018), NGC1333-IRAS4A2
(L\'opez-Sepulcre et al. 2017), Barnard 1b--S (Marcelino et al. 2018),
and L483 (Jacobsen et al. 2018).  Table 2 shows that the
CH$_3$OH/CH$_3$CHO ratio of the hot-corinos varies from $\sim$ 90 to
$\sim$ 500, that is, by less than a factor of six.  The L1157 values fall
inside this range, suggesting that the chemistry at work in the
protostellar shocks, at least in the CH$_3$OH and CH$_3$CHO context,
is similar to that ruling in the inner 100 au of the protostellar
region.  Of course, this is speculation based on one shocked region, and
 verification is required: (i) by
increasing similar measurements towards different outflow shocks, and
(ii) by investigating whether  the chemical enrichment around a protostar is
ruled by thermal evaporation or sputtering due to accretion shocks
(see e.g. Sakai et al. 2014ab, Lee et al. 2017, 2019; Codella et
al. 2018).

\begin{table}
\caption{Comparison of the CH$_3$OH/CH$_{3}$CHO abundance ratio as derived towards 
L1157-B1 with those measured towards Class 0 hot corinos using interferometric observations.}
\begin{tabular}{ccc}
\hline
\multicolumn{1}{c}{Object} &
\multicolumn{1}{c}{CH$_3$OH/CH$_3$CHO} &
\multicolumn{1}{c}{References} \\ 
\hline
L1157-B0e & 1.5 $\times$ 10$^2$ & 1 \\
L1157-B1b & 1.9 $\times$ 10$^2$ & 1 \\
HH212-mm & 0.9--3 $\times$ 10$^2$ & 2,3,4,5 \\
IRAS16293--2422B & 1 $\times$ 10$^2$ & 6,7 \\
NGC1333--IRAS4A2 & 3--6 $\times$ 10$^2$ & 8,9 \\
Barnard 1b--S & 5 $\times$ 10$^2$ & 10 \\
L483 & 2 $\times$ 10$^2$ & 11 \\
\hline
\end{tabular}

References: 1. Present work; 2. Lee et al. (2017); 3. Lee et al. (2019); 4. Bianchi et al. (2017); 
5. Codella et al. (2018); 
6. J\o{}rgensen et al. (2016); 7. J\o{}rgensen et al. 2018; 8. Taquet et al. (2015); 9. L\'opez-Sepulcre et al. (2017); 10. Marcelino et al. (2018); 11. Jacobsen et al. (2018).
\end{table}

\subsection{Chemical modelling}

The overlap between the  spatial
distribution of the CH$_3$OH and CH$_3$CHO emission suggests that either acetaldehyde or a parent molecule
was previously formed on the grain mantles, such as methanol for example, and was injected
into the gas because of the passage of the shock.  In addition, given that the
CH$_3$OH/CH$_3$CHO abundance ratio is similar in the B0 and B1 knots
despite them having a different kinematical age of less than about 1000
years, if CH$_3$CHO is synthesised in the gas phases, the process must
be relatively fast. We note that recently, Burkhardt et al. (2019) modelled 
the shocked chemistry in L1157-B1 finding that CH$_3$CHO has a 
post-shock chemistry due to sputtering, and additionally that a significant amount of its
formation is due to gas-phase processes before 10$^3$ yr.
The goal of this section is to explore whether
gas-phase reactions are indeed fast enough to be able to reproduce the
observations.

To this end, we used the same astrochemical model adopted in CC17 to
analyse the formamide and acetaldehyde emission towards L1157-B1. The
full details of the model can be found in CC17.  Briefly, we used the
time-dependent model gas-phase code MyNahoon, and adopted a two-step
procedure: (1) We first compute the chemical composition of a standard
molecular cloud at 10 K with a H$_2$ density of
  $2\times 10^4$ cm$^{-3}$ and initial elemental abundances as
  in Wakelam \& Agundez (2013) (their Table  3). (2) We then successively
increase the gas temperature, density, and gaseous abundance of grain-mantle species to simulate the shock passage. The gas temperature
  and H$_2$ density are taken as those found by the NLTE LVG
  modelling, namely 90 K and $4\times 10^5$ cm$^{-3}$,
  respectively. For the abundances of the species injected into the
  gas phase at the moment of the passage of the shock,  when
possible, we assumed abundances (with respect to H nuclei) similar to
those measured by IR observations of the interstellar dust ices
(Boogert et al. 2015): CO$_{2}$ (3 $\times$ 10$^{-5}$) and H$_{2}$O
(2 $\times$ 10$^{-4}$; see also Busquet et al. 2014).  Otherwise, we
used the values constrained by previous studies on the L1157-B1
chemistry: OCS (2 $\times$ 10$^{-6}$; Podio et al. 2014),
NH$_3$ (2 $\times$ 10$^{-5}$; Tafalla \& Bachiller 1995).  
We varied the injected CH$_3$CH$_2$ abundance in the
  gas-phase acetaldehyde formation model so as to obtain the observed
  acetaldehyde abundance. We found it necessary to inject
  CH$_3$CH$_2$/H = $ 8\times 10^{-8}$. Similarly, in the grain-surface
  acetaldehyde formation model, we modified the acetaldehyde
  abundances to fit the measured value in B1b (see \S ~4.3), namely we
  assumed CH$_3$CHO/H = $3.5\times10^{-8}$.  Finally, in both models,
  methanol is assumed to be a grain-surface product and the abundance
  injected in the gas phase is equal to that observed, namely
  $6.5\times10^{-6}$ (see \S ~4.3).

\begin{figure}
\centerline{\includegraphics[angle=90,width=12cm]{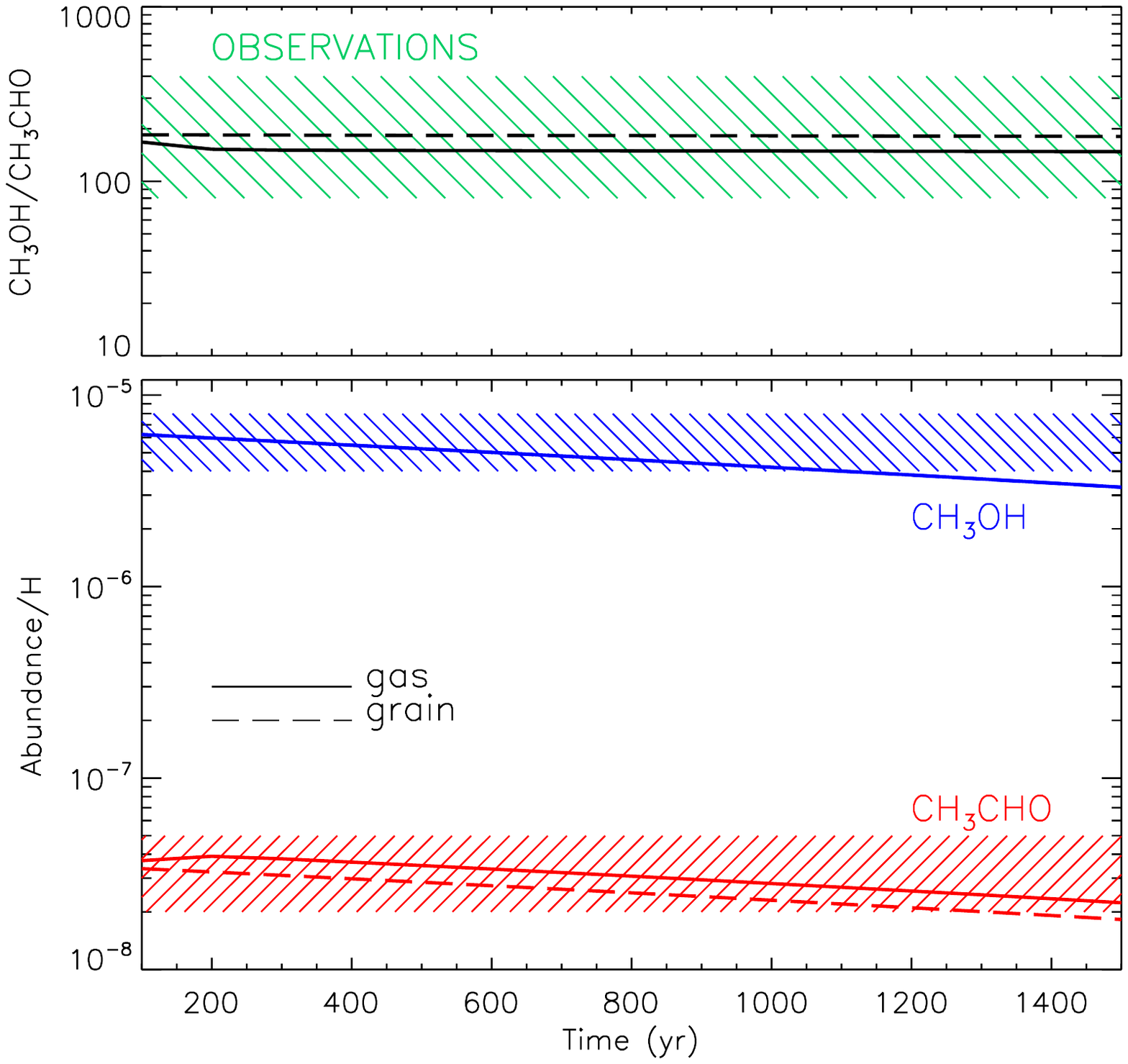}}
\caption{Model predictions of methanol and acetaldehyde after a shock
  passage.  {\it Upper panel:} CH$_3$OH/CH$_3$CHO abundance ratio:
  The solid line refers to a model where acetaldehyde is synthesised in the
  gas phase, whereas the dashed line refers to predictions assuming that
  CH$_3$OH is injected into the gas phase directly from the grain
  mantles.  The hatched green zone is defined by the value (including
  uncertainty) measured towards the B1b peak, with a kinematical
  age of $\simeq$ 1100 yr (Podio et al. 2016, revised according to the
  new distance of 352 pc by Zucker et al. 2019).  {\it Lower panel:}
  Acetaldehyde (CH$_3$CHO, red) and methanol (CH$_3$OH, blue)
  abundances with respect to H nuclei as a function of time from the
  passage of the shock.  The hatched blue and red regions show the
  measurements towards B1b (see text).}
\label{model}
\end{figure}

As in CC2017, we run both the case where acetaldehyde is directly
injected into the gas phase from the grain mantles and the case in
which it is formed in the gas phase, mainly by the reaction of ethyl
radical (CH$_3$CH$_2$) with atomic oxygen: CH$_3$CH$_2$ + O $\to$
CH$_3$CHO + H, as suggested by Charnley et al. (2004). With respect to
the chemical network used by CC2017, we included the recent
theoretical study by Gao et al. (2018) on the destruction of methanol
by the reaction with OH.

Figure 9 reports the model predictions as a function of time after the
shock passage compared with the values measured towards B1b.  As
  written above, if acetaldehyde is directly sputtered from the grain
  mantles, in order to fit the observations, its injected abundance
  must be equal to $3.5 \times10^{-8}$. On the other hand, if
  acetaldehyde is formed via gas-phase reactions, the injected ethyl
  radical abundance must be $8 \times10^{-8}$.  As shown in the
figure, the gas phase reactions are relatively fast at synthesising
acetaldehyde: within 300 years the acetaldehyde abundance reaches its
peak, after which it slowly decreases because of the destruction from ions in
the gas phase, on a timescale larger than about 2000 years.
Therefore, it is impossible to elucidate the formation route of
acetaldehyde in L1157-B1: a similar study towards an even younger shock might be able to shed light on the question.  Alternatively, a less dense  shock region, or one that was less irradiated by cosmic rays could be  good site for
better constraining the acetaldehyde formation route, as the chemical
evolution would be slowed down in those cases.

As a final remark, we would like to highlight the fact that, although
  relatively simple, the modelling described above and used in this
  work catches the basic processes at work after the passage of a
  shock and has proven to be relatively reliable for reproducing the
  observations (see e.g. Podio et al. 2014 and C2017). The most
  drastic shortcoming of our model compared to more accurate and/or sophisticated shock
  models is its lack of a higher temperature (up to 1000 K) over a
  short (a few hundred years at most) period (see e.g. Viti et
  al. 2011). This could be important in the presence of species whose
  abundances depend on reactions with activation barriers that make
  them inefficient at low ($\leq$100 K) temperatures but efficient at
  higher ones. In the specific case modelled here, we verified that no
  similar reactions are present and that a high-temperature (1000 K)
  phase does not impact our results.

\section{Conclusions}\label{sec:conclusions}

In the context of the IRAM NOEMA SOLIS Large Program we imaged six
methanol and eight acetaldehyde emission lines toward the B0 and B1
shocks along the L1157 blueshifted outflow.  The measured abundances
relative to H nuclei are $6.5\times 10^{-6}$ (methanol) and
$3.5\times 10^{-8}$ (acetaldehyde).  The CH$_3$OH and CH$_3$CHO
spatial distributions overlap well, tracing the earliest shocked
regions. Comparison with astrochemical model predictions shows that
CH$_3$CHO is either formed on the dust mantles or is quickly formed
in the gas phase using simpler mantle products.  Our
modelling demonstrates that two species emitting lines
in the same region is not enough to guarantee that they are grain-surface products: gas-phase
reactions triggered by different species from the grain mantles could
also display the same spatial behaviour. Detailed studies are required in order to discriminate between grain-surface and gas-phase chemistry as the major process leading
to the observed species and to decipher the timescale on which this formation occurs.

For acetaldehyde in particular, at present, it is not possible to
measure the abundance of one of the two parent species (ethyl radical)
possibly synthesising acetaldehyde in the gas phase. The
  measurement of the second parent species, atomic oxygen,
  is challenging, as its fine structure line at 63 $\mu$m lies in a
  spectral window that is totally blocked by the terrestrial atmosphere.  The
  SOFIA telescope could overcome this problem (Kristensen et
  al. 2017). On the other hand, although the frequencies (and
spectroscopic parameters) of ethyl radicals are  unknown at present,
they would be a very welcome target of specific spectroscopic studies.
It would then very likely be possible to search for ethyl
  radicals in regions where acetaldehyde is detected, and thus decipher whether or not
  the proposed gas-phase route is correct.

\begin{acknowledgements}
We are very grateful to all the IRAM staff, whose dedication 
allowed us to carry out the SOLIS project. 
This work was supported 
by (i) the PRIN-INAF 2016 "The Cradle of Life - GENESIS-SKA (General Conditions in Early Planetary Systems for the rise of life with SKA)", 
(ii) the program PRIN-MIUR 2015 STARS in the CAOS - Simulation Tools for
Astrochemical Reactivity and Spectroscopy in the Cyberinfrastructure
for Astrochemical Organic Species (2015F59J3R, MIUR Ministero
dell'Istruzione, dell'Universit\`a della Ricerca e della Scuola
Normale Superiore), and (iii) the European Research Council (ERC) under the
European Union's Horizon 2020 research and innovation programme, for
the Project "The Dawn of Organic Chemistry" (DOC), grant agreement No
741002.
\end{acknowledgements}

\appendix

\section{The NOEMA-SOLIS CH$_3$OH and CH$_3$CHO line maps}

Figures A.1 and A.2 show the images of the detected CH$_3$OH and CH$_3$CHO emission lines (see Table 1). 
For comparison, the CO (1--0) (white contours; Gueth et al. 1996) and 
NH$_2$CHO (4$_{\rm 1,4}$--3$_{\rm 1,3}$) (cyan contours; Codella et al. 2017) spatial distributions
are  also reported. 

\begin{figure*}
\centerline{\includegraphics[angle=90,width=18cm]{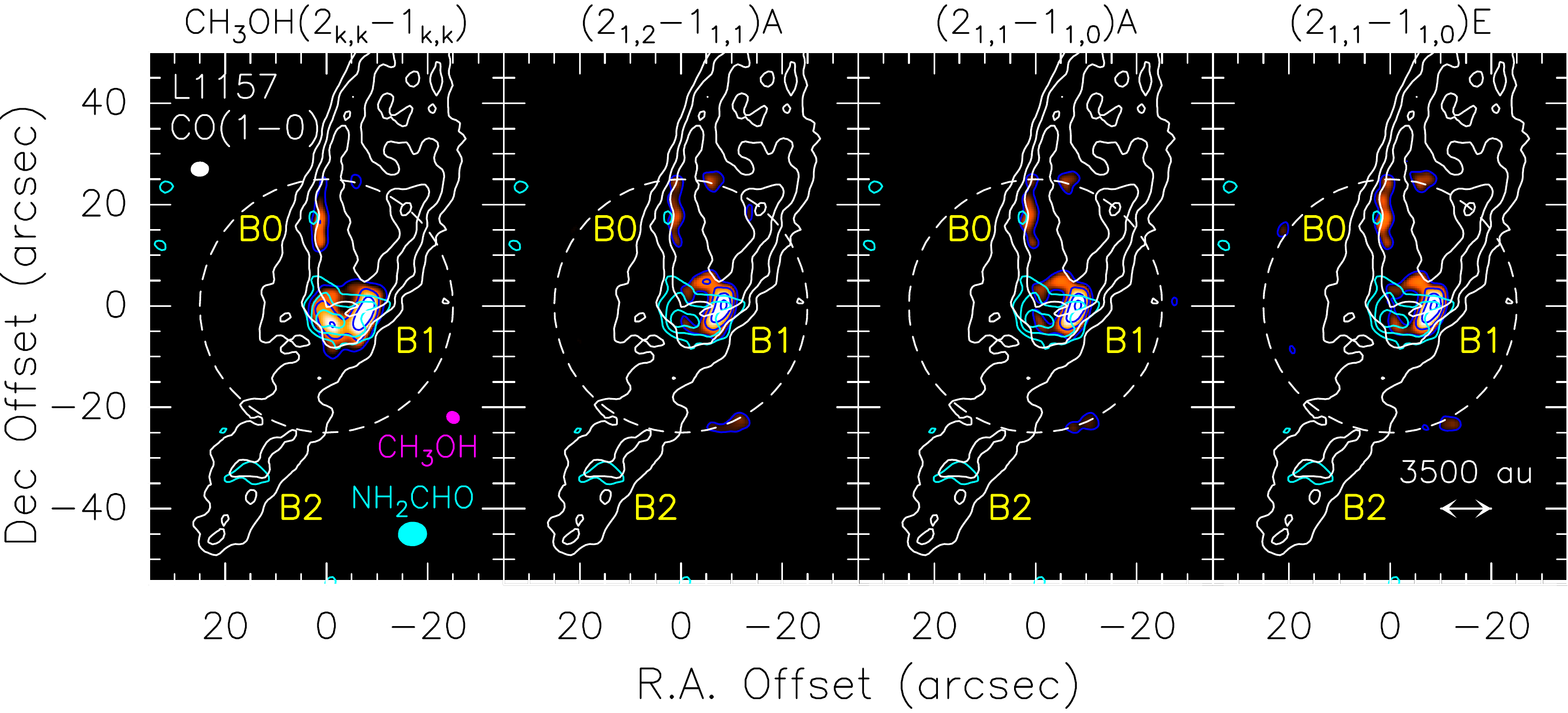}}
\vspace{-2cm}
\caption{
The L1157 southern blueshifted lobe
in CO (1--0) (white contours; Gueth et al. 1996) and  NH$_2$CHO (4$_{\rm 1,4}$--3$_{\rm 1,3}$) (cyan contours; Codella et al. 2017). 
The precessing jet
ejected by the central object L1157-mm (outside the frame and toward the northwest) excavated several clumpy cavities,
named B0,  B1, and B2, respectively. 
The maps are centred at $\alpha({\rm J2000})$ = 20$^h$ 39$^m$ 10$\fs$2,
$\delta({\rm J2000})$ = +68$\degr$ 01$\arcmin$ 10$\farcs$5 ($\Delta$$\alpha$ = +25$\arcsec$ and
$\Delta$$\delta$ = --63$\farcs$5 from the L1157-mm protostar).
Map (in colour scale) of the sum of the CH$_3$OH (2$_{\rm -1,2}$--1$_{\rm -1,1}$) E , (2$_{\rm 0,2}$--1$_{\rm 0,1}$) A,
and (2$_{\rm 0,2}$--1$_{\rm 0,1}$) E emission lines (labelled 2$_{\rm k,k}$--1$_{\rm k,k}$), (2$_{\rm 1,2}$--1$_{\rm 1,1}$) A,
(2$_{\rm 1,1}$--1$_{\rm 1,0}$) A, and (2$_{\rm 1,1}$--1$_{\rm 1,0}$) E (integrated over the whole velocity range).
For the CO image, the first contour and step are
6$\sigma$ (1$\sigma$ = 0.5 Jy beam$^{-1}$ km s$^{-1}$) and 4$\sigma$, respectively. 
The first contour and step of the NH$_2$CHO map (cyan contours) correspond to 3$\sigma$ (15 mJy beam$^{-1}$
km s$^{-1}$) and 1$\sigma$, respectively. The dashed circle shows the primary beam of the CH$_3$OH images (64$\arcsec$). 
The first contour and step are 3$\sigma$
(56 mJy beam$^{-1}$ km s$^{-1}$, 32 mJy beam$^{-1}$ km s$^{-1}$, 45 mJy beam$^{-1}$ km s$^{-1}$, and 23 mJy beam$^{-1}$ km s$^{-1}$
for  2$_{\rm k,k}$--1$_{\rm k,k}$, 2$_{\rm 1,2}$--1$_{\rm 1,1}$ A, 2$_{\rm 1,1}$--1$_{\rm 1,0}$ A, and
(2$_{\rm 1,1}$--1$_{\rm 1,0}$) E, respectively.
Yellow labels are for the B0, B1, and B2 regions: the positions of the different clumps inside these
regions are shown in Fig. 1 (see also text, and e.g. Codella et al. 2009, 2017).
The cyan, white, and magenta ellipses depict
the synthesised beams of the NH$_2$CHO (5$\farcs$79 $\times$ 4$\farcs$81, PA = --94$\degr$),
CO (3$\farcs$65 $\times$ 2$\farcs$96, PA=+88$\degr$), and CH$_3$OH (2$\farcs$97 $\times$ 2$\farcs$26, PA=--155$\degr$) 
observations, respectively.}
\label{maps_meta}
\end{figure*}

\begin{figure*}
\centerline{\includegraphics[angle=90,width=18cm]{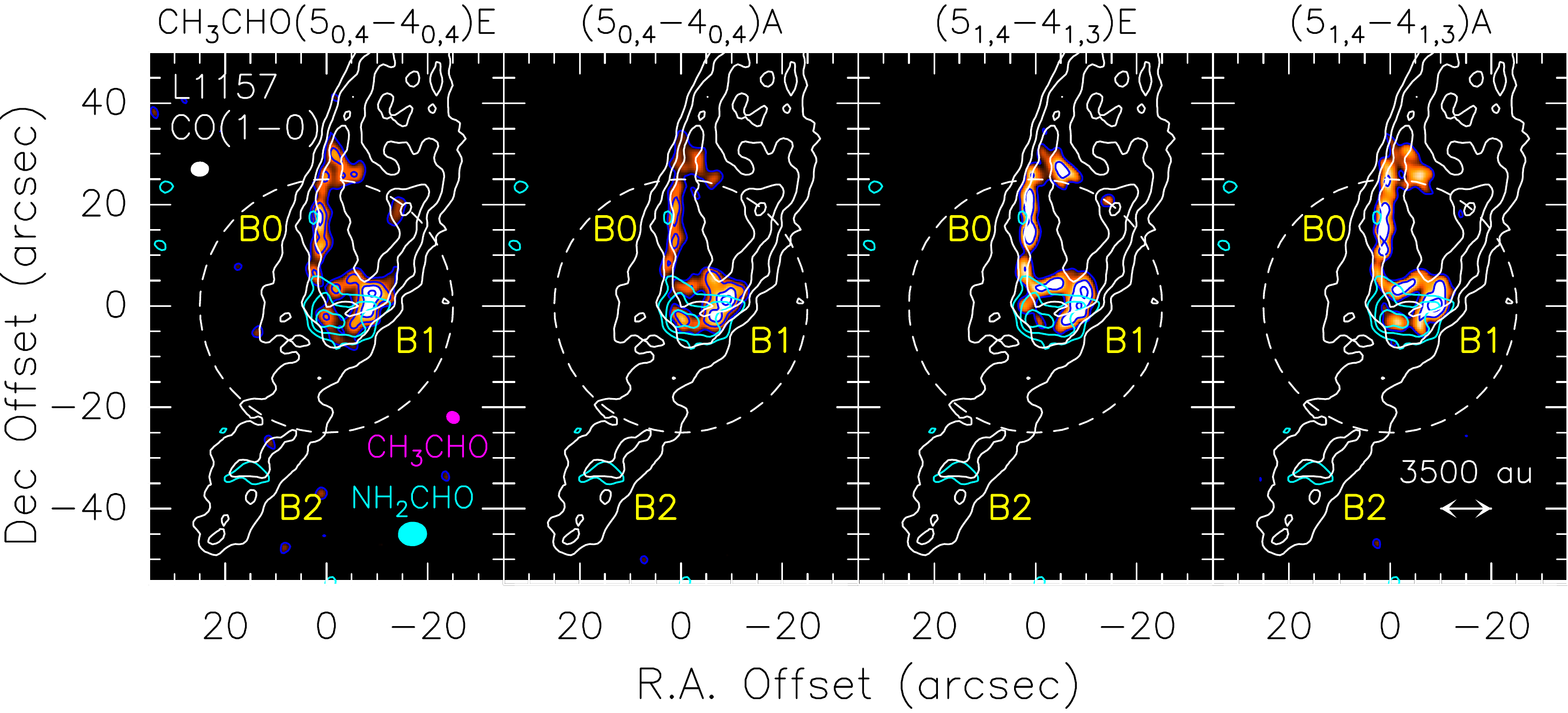}}
\vspace{-4cm}
\centerline{\includegraphics[angle=90,width=18cm]{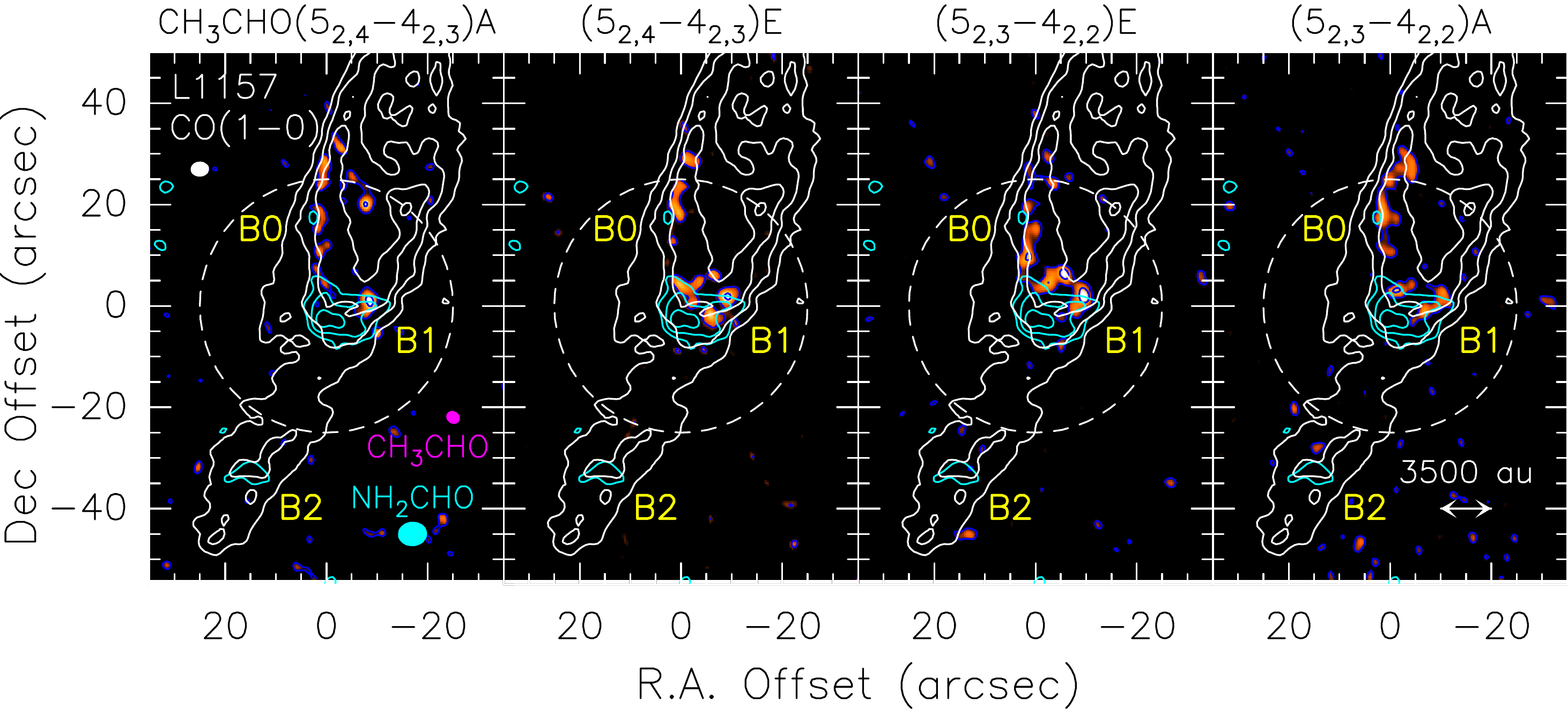}}
\vspace{-2cm}
\caption{
The L1157 southern blueshifted lobe
in CO (1--0) (white contours; Gueth et al. 1996) and  NH$_2$CHO (4$_{\rm 1,4}$--3$_{\rm 1,3}$) (cyan contours; Codella et al. 2017). 
The precessing jet
ejected by the central object L1157-mm (outside the frame and toward the northwest) excavated several clumpy cavities,
named B0,  B1, and B2, respectively. The maps are centred at $\alpha({\rm J2000})$ = 20$^h$ 39$^m$ 10$\fs$2,
$\delta({\rm J2000})$ = +68$\degr$ 01$\arcmin$ 10$\farcs$5 ($\Delta$$\alpha$ = +25$\arcsec$ and
$\Delta$$\delta$ = --63$\farcs$5 from L1157-mm).
For the CO image, the first contour and step are
6$\sigma$ (1$\sigma$ = 0.5 Jy beam$^{-1}$ km s$^{-1}$) and 4$\sigma$, respectively. 
The first contour and step of the NH$_2$CHO map (cyan contours) correspond to 3$\sigma$ (15 mJy beam$^{-1}$
km s$^{-1}$) and 1$\sigma$, respectively. The dashed circle shows the primary beam of the CH$_3$CHO images(64$\arcsec$). 
{\it Upper panels:} Map (in colour scale) of line emission (integrated over the whole velocity range) 
due to CH$_3$CHO (5$_{\rm 0,4}$--4$_{\rm 0,4}$) E, (5$_{\rm 0,4}$--4$_{\rm 0,4}$) A, 
(5$_{\rm 1,4}$--4$_{\rm 1,3}$) E, and (5$_{\rm 1,4}$--4$_{\rm 1,3}$) A, respectively. The first contour and step are 3$\sigma$:
7 mJy beam$^{-1}$ km s$^{-1}$ (5$_{\rm 0,4}$--4$_{\rm 0,4}$ E, 5$_{\rm 0,4}$--4$_{\rm 0,4}$ A), and 8 mJy beam$^{-1}$ km s$^{-1}$ 
(5$_{\rm 1,4}$--4$_{\rm 1,3}$ E, and 5$_{\rm 1,4}$--4$_{\rm 1,3}$ A). 
The cyan, white, and magenta ellipses depict
the synthesised beams of the NH$_2$CHO (5$\farcs$79 $\times$ 4$\farcs$81, PA = --94$\degr$),
CO (3$\farcs$65 $\times$ 2$\farcs$96, PA=+88$\degr$), and CH$_3$CHO (2$\farcs$97 $\times$ 2$\farcs$26, PA=--155$\degr$) 
observations, respectively).
Yellow labels are for the B0, B1, and B2 regions: the positions of the different clumps inside these
regions are shown in Fig. 1 (see also text, and e.g. Codella et al. 2009, 2017).
{\it Lower panels:} Same as {upper panels} but for the CH$_3$CHO (5$_{\rm 2,4}$--4$_{\rm 2,3}$) A, (5$_{\rm 2,4}$--4$_{\rm 2,3}$) E,
(5$_{\rm 2,3}$--5$_{\rm 2,2}$) E, and (5$_{\rm 2,3}$--4$_{\rm 2,2}$) A.
The first contour and step are 3$\sigma$:
5 mJy beam$^{-1}$ km s$^{-1}$ for 5$_{\rm 2,4}$--4$_{\rm 2,3}$ A,E, and 4 mJy beam$^{-1}$ km s$^{-1}$ for 5$_{\rm 2,3}$--5$_{\rm 2,2}$ A,E.}
\label{maps_ace}
\end{figure*}

\end{document}